\documentclass[journal]{IEEEtran}
 \usepackage{hyperref}
\ifCLASSINFOpdf
  \usepackage{subcaption}
  \usepackage[labelformat=parens,labelsep=quad, skip=3pt]{caption}
  \usepackage[pdftex]{graphicx}
  \usepackage{multicol}
  \usepackage{lipsum}
\else
  
\fi
\usepackage{siunitx}
\usepackage{graphicx}
\usepackage{booktabs}
\usepackage{multirow}
\usepackage{soul}
\usepackage{url}
\begin{document}

\title{An Uncertainty-aware Transfer Learning-based Framework for Covid-19 Diagnosis}
%
%
%

\author{
    
    Afshar~Shamsi~Jokandan,
    Hamzeh~Asgharnezhad,
    Shirin~Shamsi~Jokandan,\\
    Abbas~Khosravi,
    Parham~M.~Kebria,
    Darius~Nahavandi,
    Saeid~Nahavandi,
    and~Dipti~Srinivasan
\thanks{A. Sh. Jokandan and Sh. Sh. Jokandan are with ASR group, Tehran, Iran (e-mails: \{afshar.shamsi.j@gmail.com, shirinshamsijokandan@yahoo.com\}).}
\thanks{H. Asgharnezhad is an individual researcher, Tehran, Iran (e-mail: dara1400@gmail.com).}
\thanks{A. Khosravi, P. M. Kebria, D. Nahavandi, and S. Nahavandi are with the Institute for Intelligent Systems Research and Innovation (IISRI), Deakin University, Australia (e-mails: \{abbas.khosravi, kebria, darius.nahavandi, saeid.nahavandi\}@deakin.edu.au).}
\thanks{D. Srinivasan is with Department of Electrical and Computer Engineering, National University of Singapore (e-mail: dipti@nus.edu.sg).}
}

\maketitle

\begin{abstract}

The early and reliable detection of COVID-19 infected patients is essential to prevent and limit its outbreak. The PCR tests for COVID-19 detection are not available in many countries and also there are genuine concerns about their reliability and performance. Motivated by these shortcomings, this paper proposes a deep uncertainty-aware transfer learning framework for COVID-19 detection using medical images. Four popular convolutional neural networks (CNNs) including VGG16, ResNet50, DenseNet121, and InceptionResNetV2 are first applied to extract deep features from chest X-ray and computed tomography (CT) images. Extracted features are then processed by different machine learning and statistical modelling techniques to identify COVID-19 cases. We also calculate and report the epistemic uncertainty of classification results to identify regions where the trained models are not confident about their decisions (out of distribution problem). Comprehensive simulation results for X-ray and CT image datasets indicate that linear support vector machine and neural network models achieve the best results as measured by accuracy, sensitivity, specificity, and AUC. Also it is found that predictive uncertainty estimates are much higher for CT images compared to X-ray images.

\end{abstract}


\IEEEpeerreviewmaketitle

\section{Introduction}

\IEEEPARstart{T}{he} novel corona virus disease pneumonia (COVID-19) is a newly emerged viral disease causing a world-wide pandemic. The World Health Organization (WHO) \cite{who} has already listed the COVID-19 outbreak as the sixth international public health emergency, following H1N1 (2009), polio (2014), Ebola in West Africa (2014), Zika (2016) and Ebola in the Democratic Republic of Congo (2019). As of May 2020, it has impacted around 170 countries and regions. There are globally more than 4m identified COVID-19 cases and the number of death is fast approaching 300k. Lock downs and restrictions have been applied by authorities in different countries to slow down its spread. The impact on the world economy has been massive due to restrictions applied to people's movement and the disruption of supply chains.

Screening suspected patients and the early diagnosis of COVID-19 is the best way to prevent its outbreak within a society. The sooner the diagnosis, the faster and smoother the medical recovery.
The real-time polymerase chain reaction (real-time PCR) is the standard test for diagnosis of COVID-19 \cite{27}. There are other complementary testing frameworks as well. Chest radiography (X-ray) \cite{55} and computed tomography (CT) scanning have been used for the detection of COVID-19 \cite{37}.
These imaging techniques are more accessible in common health settings in many countries. Besides, as real-time PCR is not available at scale in many countries, the interest for COVID-19 diagnosis using medical imaging techniques has increased.

Machine learning techniques, deep learning models, and convolutional neural networks (CNNs) have been widely applied in recent years for medical imaging computer aided diagnosis \cite{Alizadeh2020, ref10, ref11, ref12, Alizadeh2019}. A deep learning framework for detection pneumonia in chest X-ray images is proposed in \cite{Jaiswal2019}. Authors use region-based CNNs to configure regional context which helps finding accurate results. \cite{Rajpurkar2017} proposes a novel deep learning algorithm that can detect pneumonia from chest X-rays at a level exceeding practicing radiologists. The CheXNet is a 121 CNN which has been trained using 112,120 frontal-view chest X-ray images individually labeled \cite{Rajpurkar2017}. 

Training multilayer CNNs requires a massive amount of data and compute resources. Currently the availability of thousands of images with proper labels is a barrier for developing reliable CNNs for the detection of COVID-19 using computer vision techniques. \cite{Ozturk2020} proposed a transfer learning-based framework for early detection of COVID-19 cases using X-ray images. The obtained accuracies for binary and multi-classes are 98.08\% and 87.02\% respectively. Authors in \cite{Apostolopoulos2020} apply the transfer learning concept and use five pretrained CNNs for extracting features and processing them using feedforward neural networks. Obtained results indicate that the VGG19 and the MobileNet outperform others in terms of the classification accuracy. Also, it is observed that the MobileNet outperforms VGG19 in terms of specificity \cite{Apostolopoulos2020}. ResNet50, InceptionV3, and InceptionResNetV2 networks are used in \cite{Narin2020} for automatic detection of COVID-19 using X-rays. Performance results suggest that the ResNet50 pretrained model achieves the highest accuracy of 98\% amongst considered CNNs.

A deep learning-based framework for COVID-19 detection using CT images was proposed in \cite{Song2020}.
The reported experimental results in the paper show that the proposed model model precisely identifies the COVID-19 cases from others with an AUC of 0.99 and recall (sensitivity) of 0.93. In \cite{Barstugan2020}, machine learning techniques are applied for detection of COVID-19 cases from patches obtained from 150 CT images. Authors in \cite{Zheng2020} demonstrate that weakly-supervised deep learning algorithms could achieve promising results for COVID-19 detection. The number of collected sample is 499 which are processed using segmentation techniques and 3D CNNs. A deep learning framework is also proposed in \cite{Xu2020} for detection COVID-19 and influenza-A viral pneumonia. The overall accuracy of developed models for 618 CT images is 86.7\%.

All these studies report promising results for CNN models trained using a limited number of images. Deep neural networks often have hundred and thousands of trainable parameters that their fine tuning requires massive amounts of data. Besides, the limited number of samples raise concerns about epistemic uncertainty \cite{Postels2019}. It is not clear how one can trust these models for a new case assuming they have been developed using a very limited number of training samples. These models could easily fail in real world applications if the training and testing samples are different \cite{Hendrycks2016} or far from the support of the training set (out of distribution samples) \cite{liu2020simple}. None of these models are able to report their lack of confidence for new cases. That information is essential for their widespread deployment as a reliable medical diagnosis tool \cite{Amodei2016}. Identifying and flagging these difficult to predict samples has much more practical values than correctly classifying them. A radiologist may consult with their senior colleagues when dealing with ambiguous or unknown cases. Accordingly, it is too important for DNNs to generate uncertainties as an additional insight to their point estimates \cite{ghoshal2020estimating}. This extra insight greatly improves the overall reliability in decision-making as the user will know when and where they can trust predictions generated by the model. The unflagged erroneous diagnosis could lead to unfortunate life losses which could easily blockade further machine and deep learning applications in medicine \cite{Postels2019}. 

Motivated by these shortcomings, this paper proposes a novel transfer learning-based and uncertainty-aware framework for reliable detection of COVID-19 cases from X-ray and CT images. We use four pretrained CNN models (VGG16, DenseNet121, InceptinResNetV2 and ResNet50) to hierarchically extract informative and discriminative features from X-ray and CT images. This transfer learning approach is essential and efficient considering the limited number of samples. Extracted deep features are then passed to a number of machine learning model for the supervised classification task. Different performance metrics are computed for the comprehensive evaluation and the fair comparison of obtained results from different CNN architectures and classifiers. Last but not least, we also investigate the impact of lack of data on the reliability and quality of the classification results. The type of uncertainty that is important for deep learning models used for COVID-19 diagnosis is epistemic uncertainty which captures the model lack of knowledge about the data \cite{Amersfoort2020}. We then develop an ensemble of neural network models trained using different deep features to generate predictive uncertainty estimates. The quantified epistemic uncertainties provide informative hints about where and how much one can trust the model predictions.

The rest of the paper is organized as follows. Section \ref{Sec:Proposed Method} introduces our proposed method for classification and uncertainty quantification. Datasets and classification techniques are explained and introduced in section \ref{Sec:ExpSetup}. Section \ref{Sec:simulations} discusses obtained results and simulations in detail. Finally, the study is summarized in section \ref{Sec:concl}.

\section{Proposed Method}\label{Sec:Proposed Method}

\subsection{Transfer Learning-based Classification}
We will here apply the transfer learning approach to train machine learning models for COVID-19 detection. Two major issues motivate us to solve the COVID-19 detection using a transfer learning framework: (i) training DNN/CNN models require massive amount of data. This is not practical for COVID-19 as the number of collected and labeled images is very limited and often in the order of a few hundreds; (ii) training DNN/CNN models is computationally demanding. Even if thousands and millions of images are available, still it makes sense to first check the usefulness of existing pretrained models for data representation and feature extraction.

The proposed framework purely uses information content of X-rays and CT images to identify the presence of COVID-19. Here we consider five pretrained networks on ImageNet dataset and import and adapt them for the task of COVID-19 detection. These networks are VGG16 \cite{VGG2014}, ResNet50 \cite{ResNet2015}, DenseNet121, and InceptionResNetV2 \cite{goingdeep}. All these networks have achieved state of the art performance for correctly classifying images of the ImageNet dataset. Training of these networks is computationally very demanding as they have many layers and millions of trainable parameters. The main hypothesis in the proposed framework is that there are fundamental similarities between image detection/recognition tasks and the binary classification problem of COVID-19 using images. Accordingly, learnings from the former one can be safely ported to the later one to shorten the training process. While all five pretrained networks have been developed using non-medical images, it is reasonable to assume that their transformation of X-ray and CT image pixels could make the classification task easier. 

\begin{figure}[t]
\centering
\includegraphics[width=0.48\textwidth]{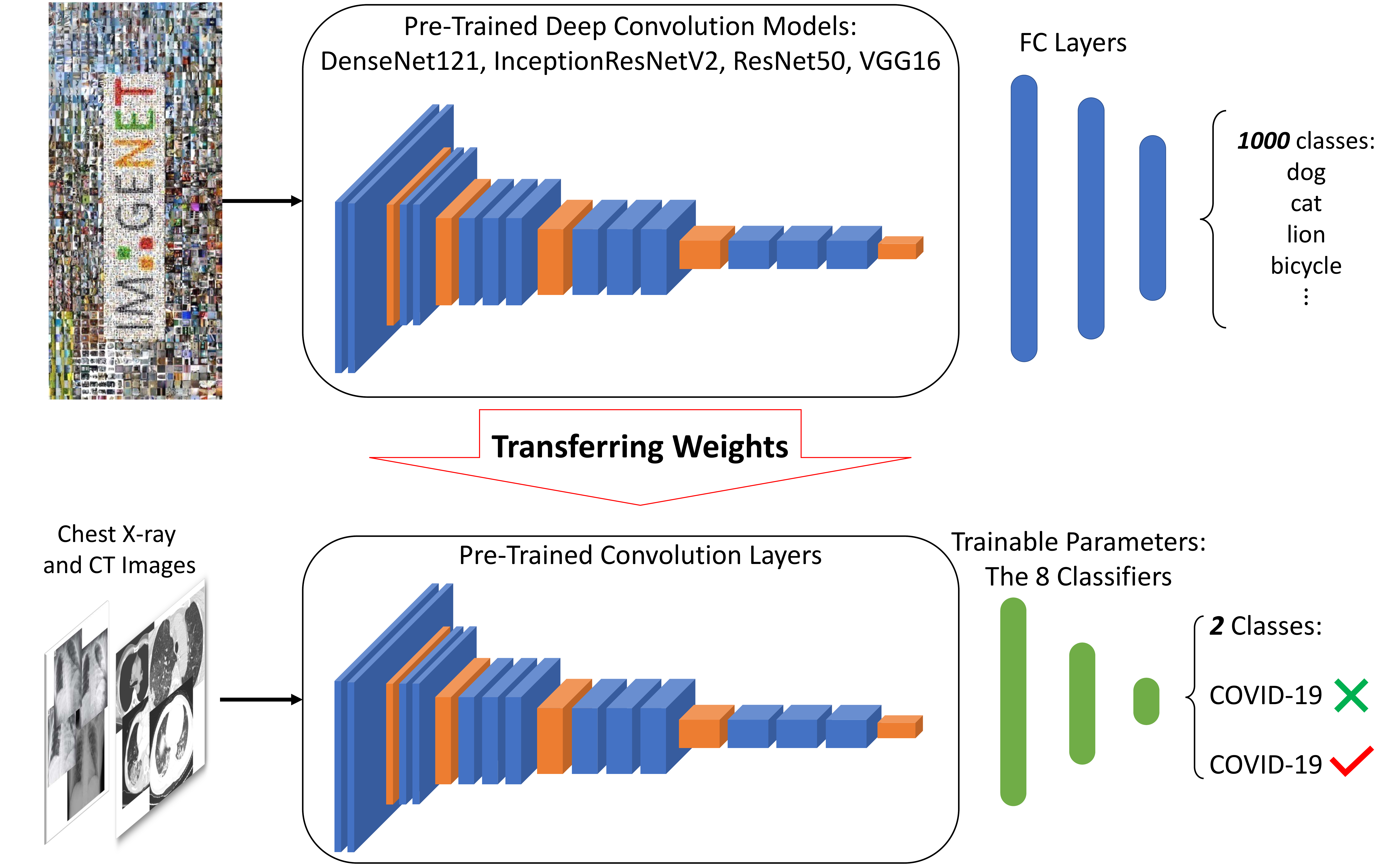}
\caption{Block diagram of the proposed transfer learning-based framework for the COVID-19 detection using X-ray and CT images.}
\label{Fig:architecture}
\end{figure}

As shown in Fig. \ref{Fig:architecture}, the parameters of the convolutional layers are kept frozen during the training process. The convolutional layers of these five pretrained models are fed by X-ray and CT images for hierarchical feature extractions. The front end of the pretrained networks is then replaced by different machine learning classifiers to separate Covid and non-Covid cases. It is important to mention that we drop the pooling operation in the last convolutional layer of these pretrained networks. This is to avoid loosing informative features before passing them to the classification models.

\subsection{Uncertainty Quantification}
Any classification study without reporting the predictive uncertainty estimates is not complete. 
There are two types of uncertainties which needs to be considered for deep learning models \cite{Kendall2017}: 

\begin{itemize}
  \item Aleatoric uncertainty which is related to the noise inherent in the data generating process. This type of uncertainty is irreducible. 
  
  \item Epistemic uncertainty which captures the ignorance about the model. In contrast to aleatoric uncertainty, epistemic uncertainty is reducible with collection of more training samples from diverse scenarios \cite{Alizadehsani2020UQ}.
  
\end{itemize}

In this paper, we mainly focus on epistemic uncertainty as it closely relates to the generalization power of models for new samples \cite{Lakshminarayanan2016} \cite{Kabir2019} \cite{Quan2019}. Here we will use an ensemble of diverse models to obtain uncertainties associated with made inferences \cite{Lakshminarayanan2016}. An ensemble consists of several models developed with different architectures, types, and sampled subsets. These model development differences cause diversity in the generalization power of models. Predictions obtained from individual models are then aggregated to obtain the final prediction. The prediction variance could be used for the calculation of the epistemic uncertainty \cite{Hosen2014}.

Similar to work in  \cite{Amersfoort2020} and \cite{Gal}, we calculate the prediction entropy as a measure of the epistemic uncertainty. the prediction entropy is a metric to measure the uncertainty in scores generated by different models \cite{Gal}. The ensemble epistemic uncertainty is calculated as the entropy of the mean predictive distribution (by averaging all predicted distribution):

\begin{equation}
    \hat{p} (y|x)\ = \ \frac{1}{N}\ \sum\limits_{i=1}^N p_{\theta_{i}} (y|x) \label{eq:1}
\end{equation}

\begin{equation}
    H (\hat{p} (y|x)) \ = \ \sum\limits_{i=0}^C \hat{p}(y_i | x)\ log\ \hat{p}(y_i | x) \label{eq:2}
\end{equation}

\noindent where $\theta_i$ is the set of parameters for $i_{th}$ network element, and $C$ ranges over all classes. For instance, suppose for a given input, an individual neural network predicts that the input is belongs to class 1 with x amount of probability and to class 0 with y amount. If we repeat this procedure 10 times for that specified input, it is similar to ensembling 10 networks for predicting the output probability. The final output probability can be calculated using Eq.~\ref{eq:1}. Now, imagine the average probability predicts that an input belongs to class 1 and 0 with 0.6 and 0.4 respectively. Based on Eq.~\ref{eq:2}, the prediction entropy can be calculated as $0.6 * log(0.6) + 0.4 * log(0.4) $. It is obvious that the prediction entropy is become zero when the output is assigned to a class with high probability, and become maximum when the network is uncertain about its outcome.

\section{Experiment Setup}\label{Sec:ExpSetup}

\subsection{Datasets}\label{Sec:Dataset}
There are two types of datasets used in this study: chest X-ray and breast CT scan. These two types of imagery datasets are the main sources of information that clinicians use for COVID-19 diagnosis. The description of these datasets is provided in this section. Also statistical and machine learning classifiers applied to process features extracted by CNNs are briefly introduced.

\subsubsection{Chest X-ray Dataset}
\begin{figure}[t]
   \centering
   \begin{subfigure}{.45\columnwidth}
      \includegraphics[width=4cm,height=2.75cm]{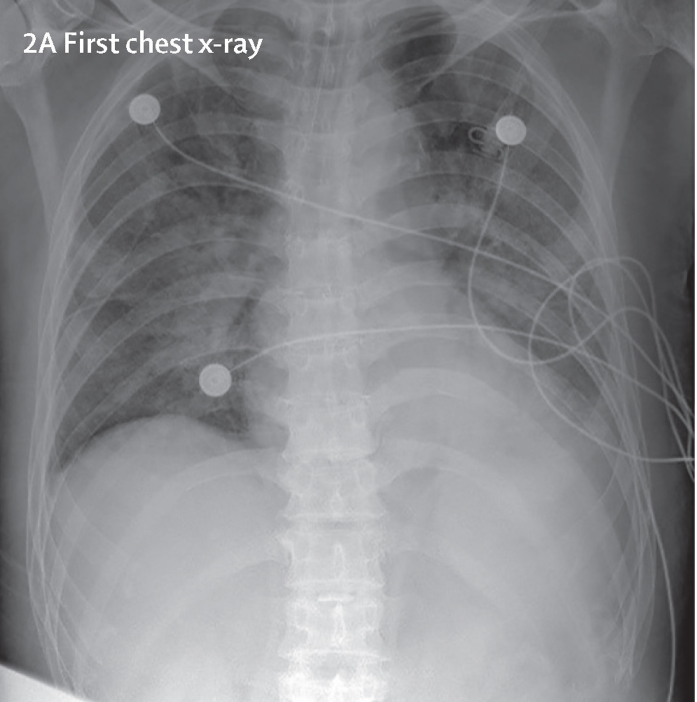}
      \caption{Covid}
      \label{fig:x-ray-covid}
  \end{subfigure}
  \begin{subfigure}{.45\columnwidth}
    \centering\includegraphics[width=4cm,height=2.75cm]{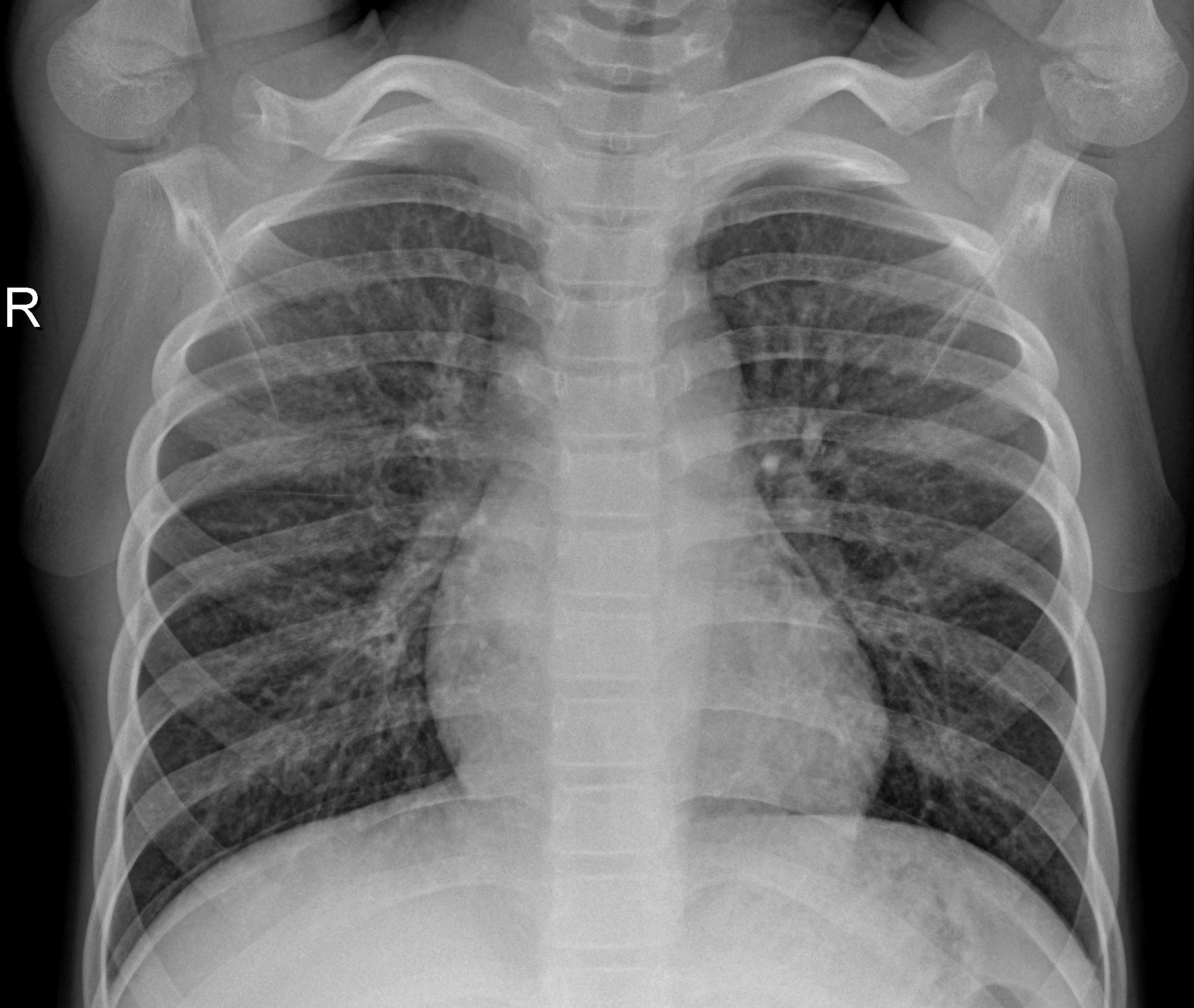}
    \caption{Normal}
    \label{fig:x-ray-normal}
  \end{subfigure}
  \caption{Two sample images from X-ray dataset.}
  \label{fig:x-ray}
\end{figure}

This dataset is formed by taking 25 images of COVID-19 from \cite{cohen2020covid} in the first step. We then add another 75 non-Covid cases of chest X-ray image from \cite{Pneumonia}. It is important to note that these non-Covid (normal) cases might consist of other unhealthy conditions such as bacterial or viral infections, chronic obstructive pulmonary disease and even a combination of two or more. Accordingly, what we mean by a normal or non-Covid case does not necessarily infer a healthy lower respiratory system. Two images of covid and normal classes are shown in Fig. \ref{fig:x-ray}. Fig. \ref{fig:x-ray-normal} displays a normal (non- Covid) case, yet virally infected. All images in this dataset are accessible via this link: https://github.com/dara1400/Covid19-Xray-Dataset.

\subsubsection{CT Dataset}
\begin{figure}
\centering
\begin{subfigure}{.45\columnwidth}
    \includegraphics[width=\columnwidth]{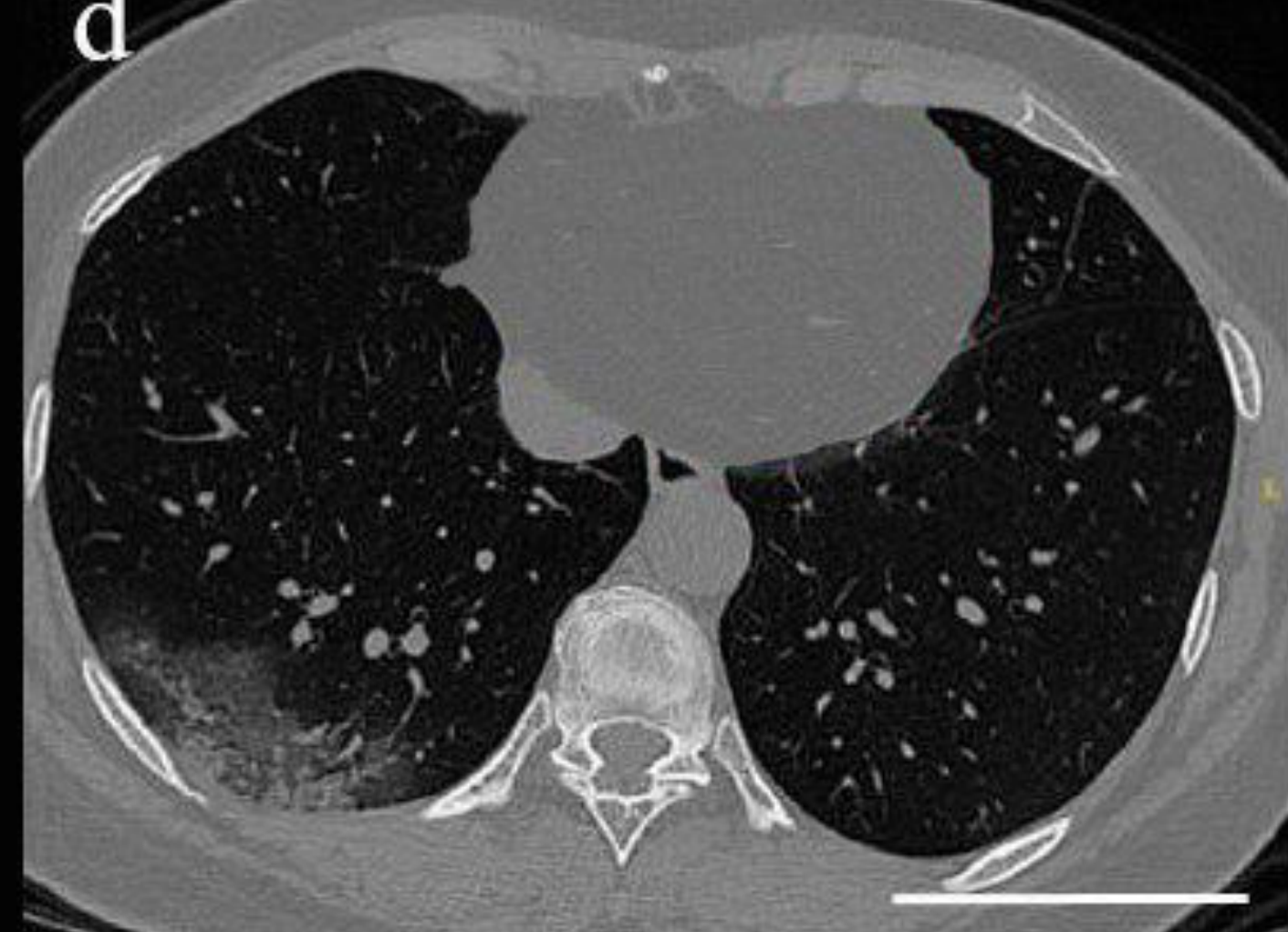}
    \caption{Covid}
  \end{subfigure}
  \begin{subfigure}{.45\columnwidth}
    \includegraphics[width=\columnwidth]{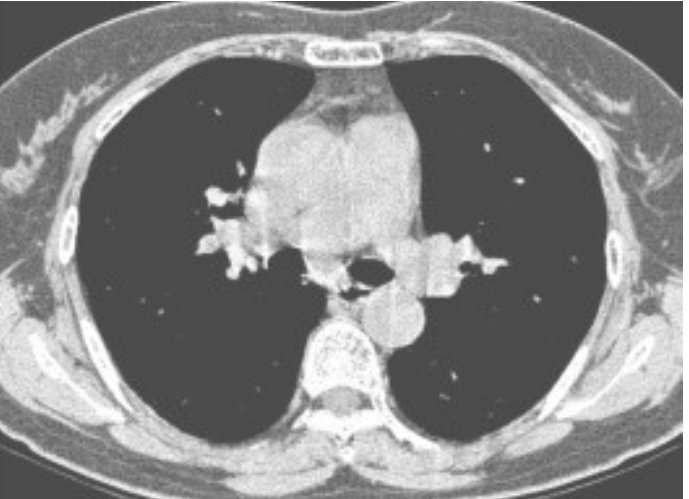}
    \caption{Normal}
    
  \end{subfigure}
  \caption{Two sample images from CT dataset.}
  \label{fig:ct}
\end{figure}

CT dataset has 349 Covid images and 397 non-Covid images \cite{60}. Health professionals prefer breast CT scans as they carry more information compared to chest X-rays to use for medical diagnosis. Fig. \ref{fig:ct} shows both a Covid and a non-Covid case from the CT database.

\subsection{Pretrained Models}
Here we briefly introduce the four CNNs used in this study for extracting features. 

\subsubsection{VGG16} \cite{VGG2014}
This model is similar to AlexNet and consists of $13$ convolution, nonlinear rectification, pooling and $3$ fully connected layers~\cite{khan34}. The filter size of convolutional network is $3*3$ and the pooling size is $2*2$. Due to its simple architecture, VGG network is performed better than AlexNet.

\subsubsection{ResNet50} \cite{ResNet2015}
Residual convolutional network (ResNet) is one the most popular deep structure which is used for classification problem (winner of ImageNet competition in 2015). Residual blocks enables the network to provide direct path to its early layers. This helps the gradient flow easily in the back propagation algorithm.

\subsubsection{DenseNet121} \cite{DenseNet2017}
DenseNet won the ImageNet competition in 2017. Traditional deep networks have only one connection between layers. However in DenseNet, all layers receive all feature maps from previous layers as input \cite{DenseNet2017}. This helps the network to decrease the number of parameters and also relieve the gradient vanishing.

\subsubsection{InceptionResNetV2}
Szegedy et al \cite{goingdeep} presented a novel structure which helps to go deeper through convolution networks. Deep networks are prone to overfitting. They solve this solution using inception blocks. Further, they use residual blocks and create InceptionResNetV2 which use the combination of residual and inception blocks wisely.

High level information about these pretrained models is provided in Table \ref{Tab:cnnInfo}. As illustrated in Fig. \ref{Fig:architecture}, network weights are kept frozen during the transfer learning procedure. The size of our input images is 224$\times$224 for VGG16, ResNet50 and DenseNet121. The input size for the InceptionResNetV2 architecture is 299$\times$299. 

\begin{table*}[t]
\caption{Information about four considered architectures for transfer learning} \label{Tab:cnnInfo}
 \centering
\begin{tabular}{lccccc}
\hline

Architecture & Paper & Year Proposed & Input Size & Number of Features & Number of Parameters \\ \hline

VGG16 & 2014 & \cite{VGG2014} & 224$\times$224 & 25,088 & 14,714,688 \\

ResNet50 & 2015 & \cite{ResNet2015} & 224$\times$224 & 100,352 & 23,587,712 \\

DenseNet121 & 2017 & \cite{DenseNet2017} & 224$\times$224 & 50,176 & 7,037,504\\

InceptionResNetV2 & 2015 & \cite{goingdeep} & 299$\times$299 & 98,304 & 54,336,736\\ [1ex]

\hline

\end{tabular}
\end{table*}

\subsection{Classification Methods}
The COVID-19 detection is a binary classification problem where the input is an image (chest X-ray or CT image) and the output is a binary label representing the presence or absence of COVID-19. 
Here images are first processed by the convolutional layers of five pretrained networks. Hierarchically extracted features are then processed by multiple classifiers. We use eight classifiers to process features: k-nearest neighbors (NN), linear support vector machine (linear SVM), radial basis function (RBF) SVM, Gaussian process (GP), random forest (RF), multi-layer perceptron (NN), Adaboost, and Naive Bayes. These classifiers are briefly introduced below:

\subsubsection{k-Nearest Neighbor (kNN)}
kNN is one of the simplest classification algorithm. It keeps a copy of all samples and classifies samples based on a similarity measure. This similarity measure is usually a kind of distance in the feature space. Most commonly used distance measures are Euclidean and Minkowski. In this paper, we use $k=2$ and Minkowski distance metric for the classification task. The Minkowski is calculated as this:

\begin{equation}\label{Eq:Minkowski}
        D(x, y) = \left( \sum_{i=1}^n \left | x_i - y_i \right |^p \right)^{1/p}
\end{equation}

If p =2, the Minkowski distance is the same as the Euclidean distance.
    
\subsubsection{Support Vector Machine (SVM)}
It is a practical solution for classification problem especially in high dimensions. SVM uses line or hyperplane for dividing the data into appropriate classes. It tries to find a hyperplane with the largest distance to the most near data for each class (margin). The lower generalization error will be achieved when the margin becomes large \cite{bishop}.
    
\subsubsection{RBF SVM}
It is a kind of SVM which uses the radial basis function (RBF) kernel for calculating the similarity (distance) between two samples. RBF kernel for two typical samples ($x, y$) can be calculated as below:
    
\begin{equation}
    K(x, y)\ =\ \exp(- \frac{||x - y||}{2\sigma^2})  
\end{equation}

\subsubsection{Gaussian Process (GP)}
It is a set of random variables in a way that each set is described by a multivariate normal distribution. The final distribution of a GP is a joint distribution of all those random variables. GP uses covariance matrix and its inversion thus it will be a lazy learning algorithm in high dimension space. It outputs a distribution which not only estimates the prediction, but also provides prediction uncertainty estimates. We use RBF kernel with a length-scale equal to one for GP classifiers in this study.

\subsubsection{Neural Network (NN)}
A feedforward NN finds a nonlinear mapping between the fixed size inputs and the output (target). The network is composed of several hidden layers and processing units called neurons. A neuron receives inputs from neurons of the previous layer and generates its own output based on the assigned activation function. The connection weights between layers of the network are trained using training algorithms such as stochastic gradient descent or adaptive moment estimation (Adam). 

\subsubsection{Random forest (RF)}
RF classifier includes several decision trees developed in parallel. These trees are developed by randomly selecting a subset of features and samples from the original training samples. Each tree will vote and the class which has the most votes will be the final prediction. In this paper, we set the number of decision trees to 10.

\subsubsection{Adaboost}
Adaboost forms an efficient classifier by mixing several weak classifiers. Classifiers are formed in a serial approach in contrast to RF where classifiers are formed in parallel. Each classifier focuses on fixing mistakes made by previous classifiers. We here set the number of weak classifiers to 50.
 
\subsubsection{Naive Bayes}
Naive Bayes classifiers are the simplest Bayesian networks that use Bayes theorem. We use Gaussian naive Bayes which predict a posterior using a normal prior based on the Bayes theory.    

Full information about these classifiers could be found in basic machine learning and statistical books \cite{james2013, bishop2006pattern, hastie2009elements, murphy2012}.

\section{Simulations and Results}\label{Sec:simulations}
The simulation results and discussions are provided in this section. We first present results obtained by different classifiers processing features extracted by pretrained CNNs. We then focus on the uncertainty quantification problem using NN models and discuss its importance for the diagnosis of COVID-19.
 
\begin{figure}[!t]
\begin{subfigure}{4.2cm}
    \centering
    \includegraphics[width=4cm]{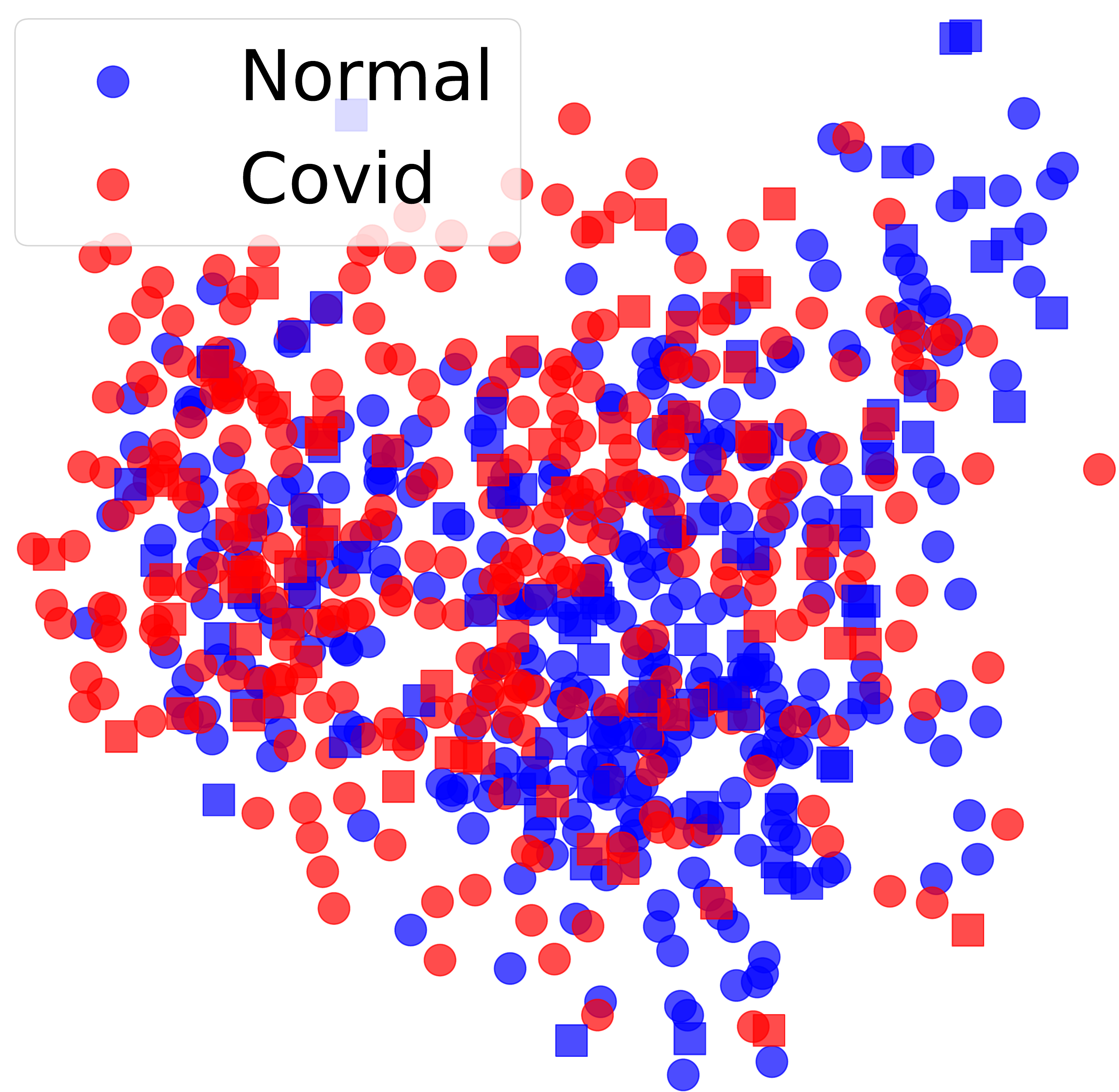}
    \caption{CT}
  \end{subfigure}
  \begin{subfigure}{4.2cm}
    \centering
    \includegraphics[width=4cm]{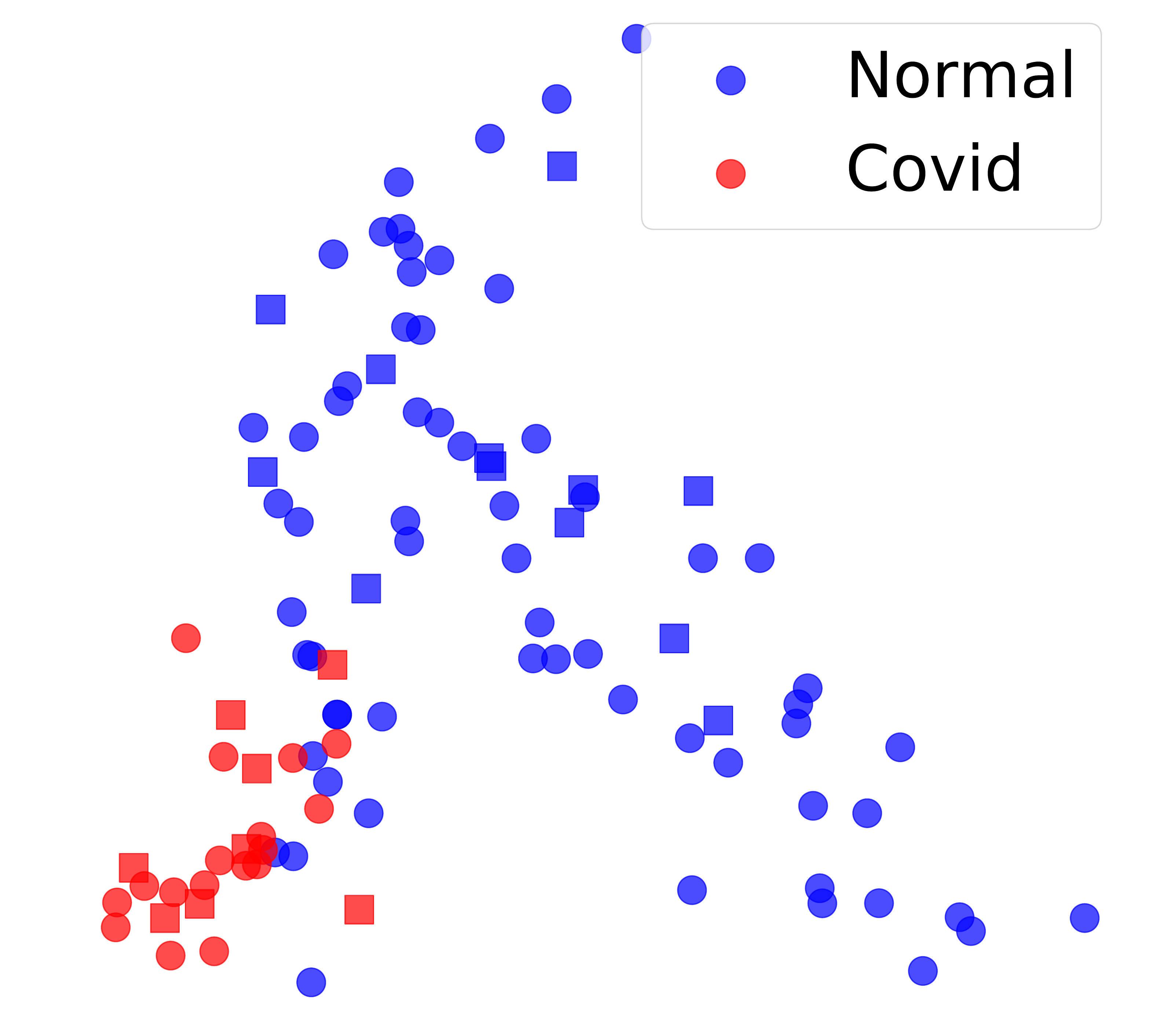}
    \caption{X-ray}
  \end{subfigure}
  \caption{2D representation of X-ray and CT images processed by VGG16 and principal component analysis.}
  \label{Fig:PCA2}
\end{figure}

To build an intuition about samples, we first extract deep features using the method shown in Fig. \ref{Fig:architecture} and then map them to the 2D space using principal component analysis (PCA) algorithm. Fig. \ref{Fig:PCA2} displays VGG16 features in the 2D space. As reported in Table \ref{Tab:cnnInfo}, the total number of extracted features for VGG16 is 25,088. Obviously, the number of samples for CT dataset is much higher and it is more balanced than the X-ray dataset. Also, it is interesting to see that normal (non-Covid) and Covid cases are fairly distinguishable for X-ray images in 2D space. In contrast, the decision boundary cannot be reasonably drawn in two dimensions for CT images. 

\begin{figure}
	\centering
	\begin{subfigure}{\linewidth}
		\includegraphics[width=\linewidth]{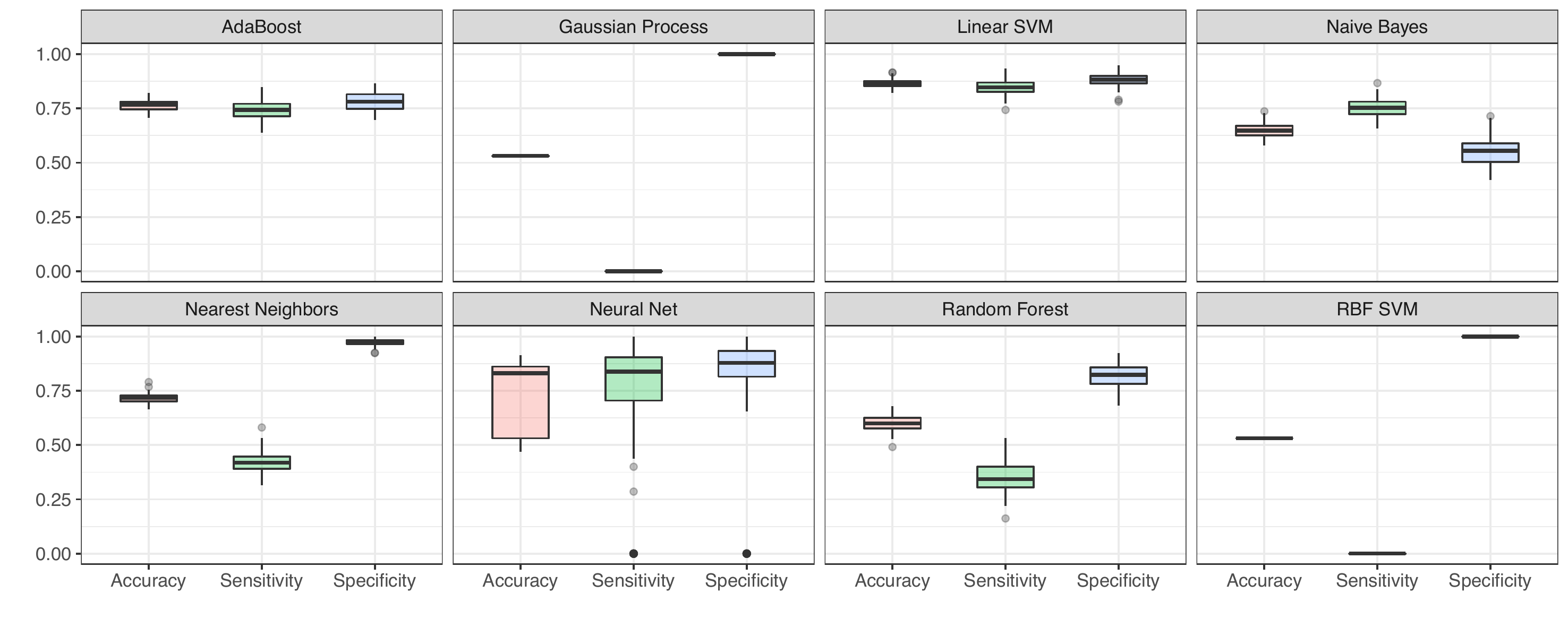}
		\caption{CT}
		\label{subfig1}
		
     \end{subfigure}
     \begin{subfigure}{\linewidth}
	  \centering
		\includegraphics[width = \linewidth]{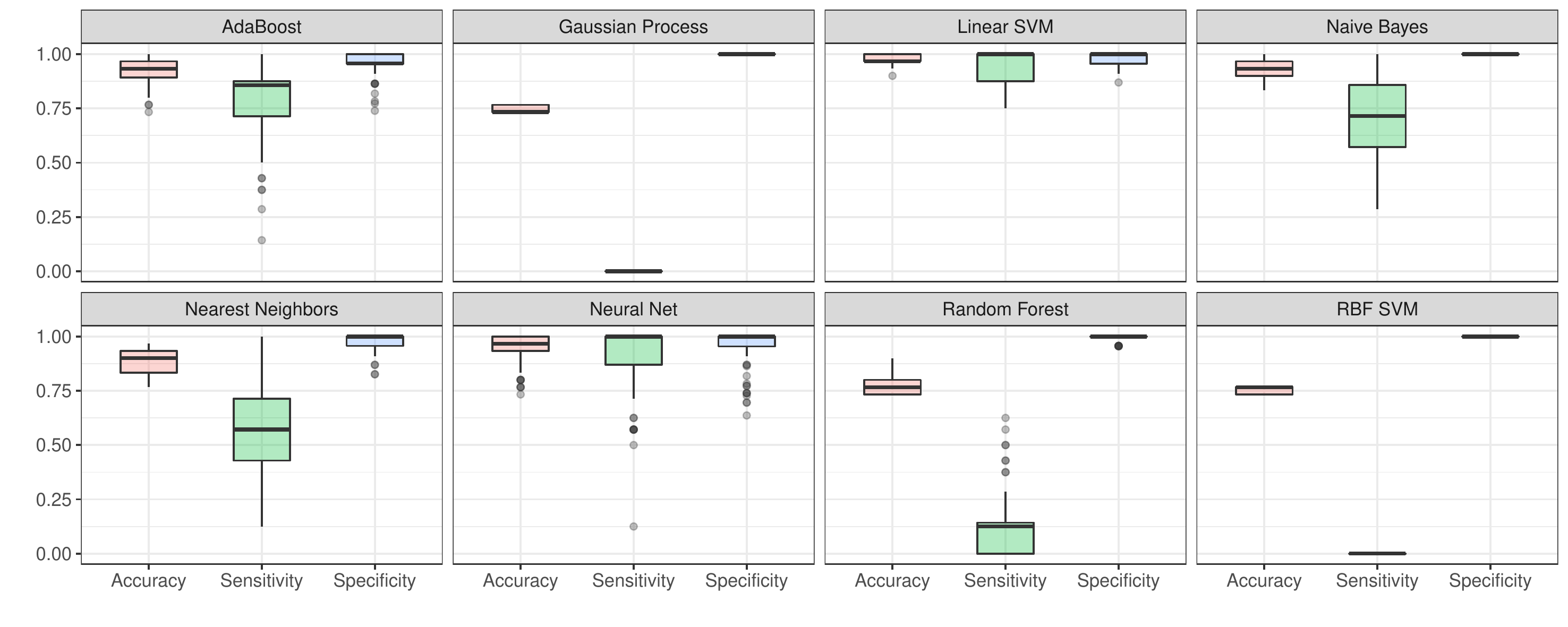}
		\caption{X-ray}
		\label{Comp1Force}
		
		\end{subfigure}
	\caption{The figures represent the distribution of Accuracy, Sensitivity and Specificity associated with CT and X-ray datasets, respectively (the top results are for CT dataset of our different classifiers).}
	\label{Fig:Boxplot}
\end{figure}

Accuracy, sensitivity, and specificity are considered for the model evaluation. Purely relying on accuracy could lead to misleading results as both datasets and in particular X-ray one are imbalanced. For obtaining statistically valid conclusions, we train every single classifier 100 times using obtained features from pretrained CNNs. For each run, the performance metrics are calculated and then the box plot graph is generated. Fig. \ref{Fig:Boxplot} represents the box plots for CT and X-ray datasets, respectively for the accuracy, sensitivity and specificity. It is noted that those values are calculated without PCA for all classifiers trained 100 times (all features passed to classifiers). 

It is observed that the more the number of features, the less the ability of RBF SVM and GP to correctly classify samples. This is because they use the covariance matrix and its inversion, which in turn drops the sensitivity value to zero, making them unreliable. In contrast, linear SVM and NN prove to be the best ones among considered classifiers. This is because they use simple hyperplanes to separate features of two classes.
    
\begin{figure*}
\centering
\begin{subfigure}[b]{0.24\textwidth}
\includegraphics[width=\textwidth]{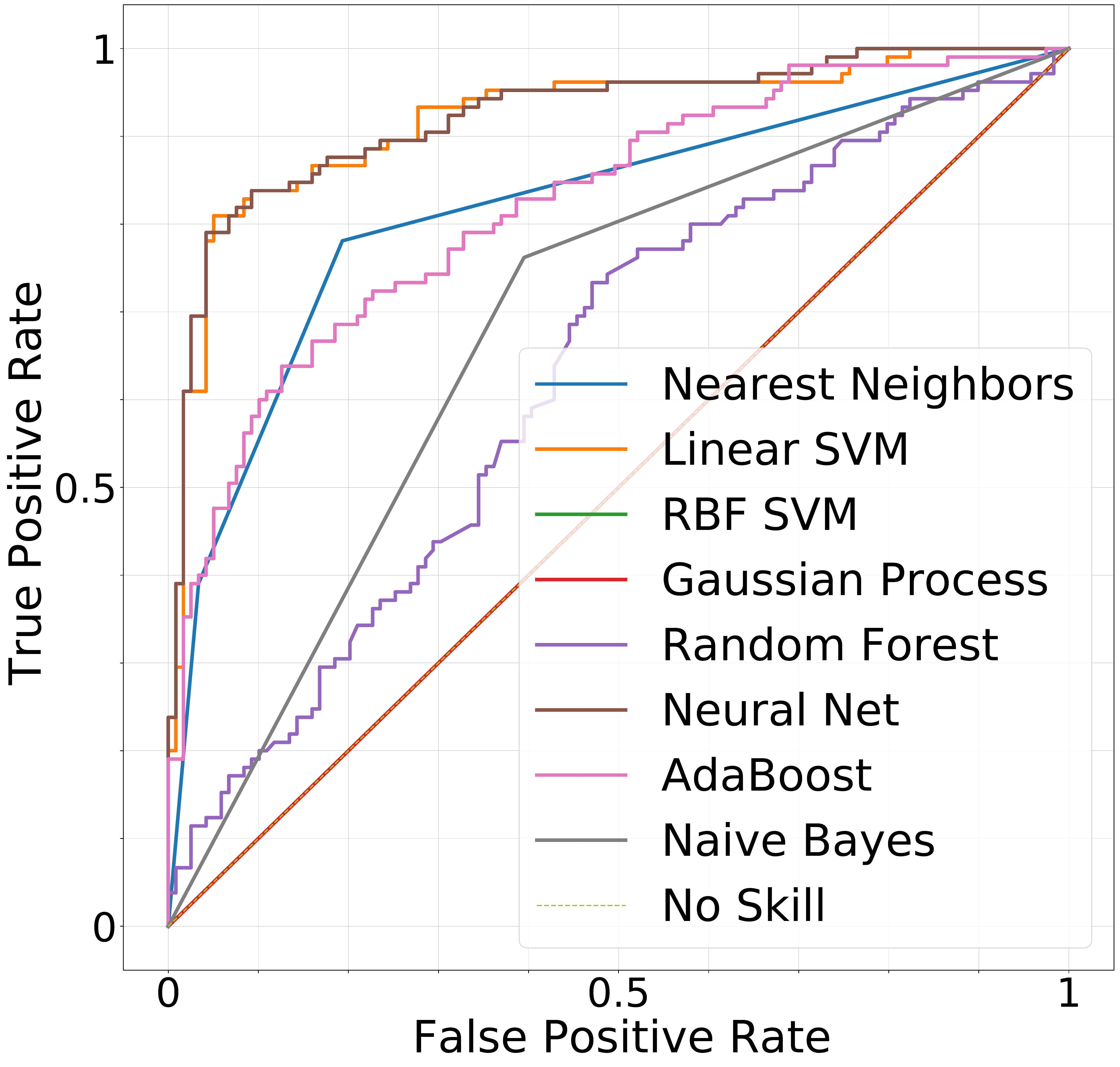}
\caption{VGG16}
\end{subfigure}
\begin{subfigure}[b]{0.24\textwidth}
\includegraphics[width=\textwidth]{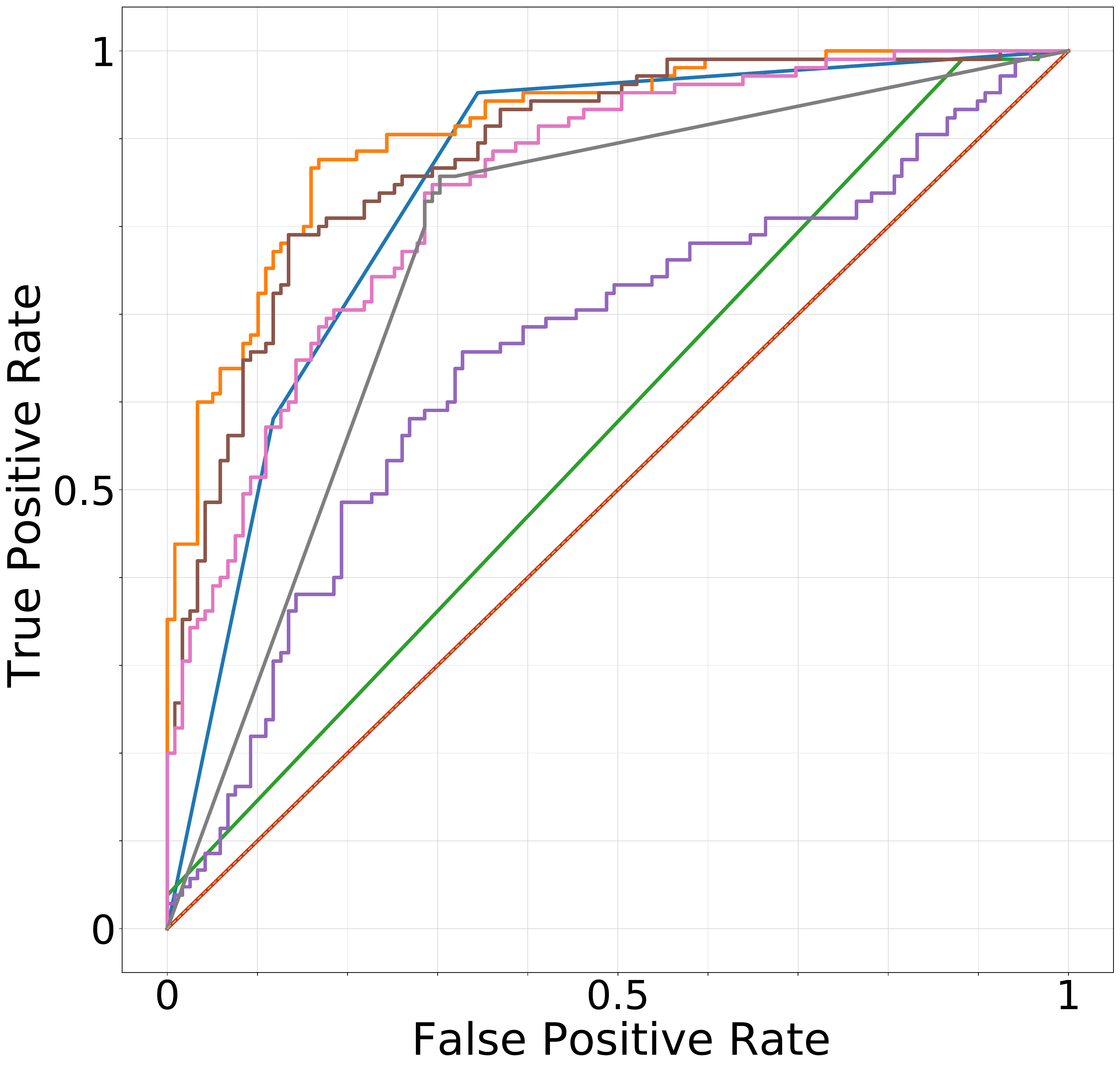}
\caption{InceptionResNetV2}
\end{subfigure}
\begin{subfigure}[b]{0.24\textwidth}
\includegraphics[width=\textwidth]{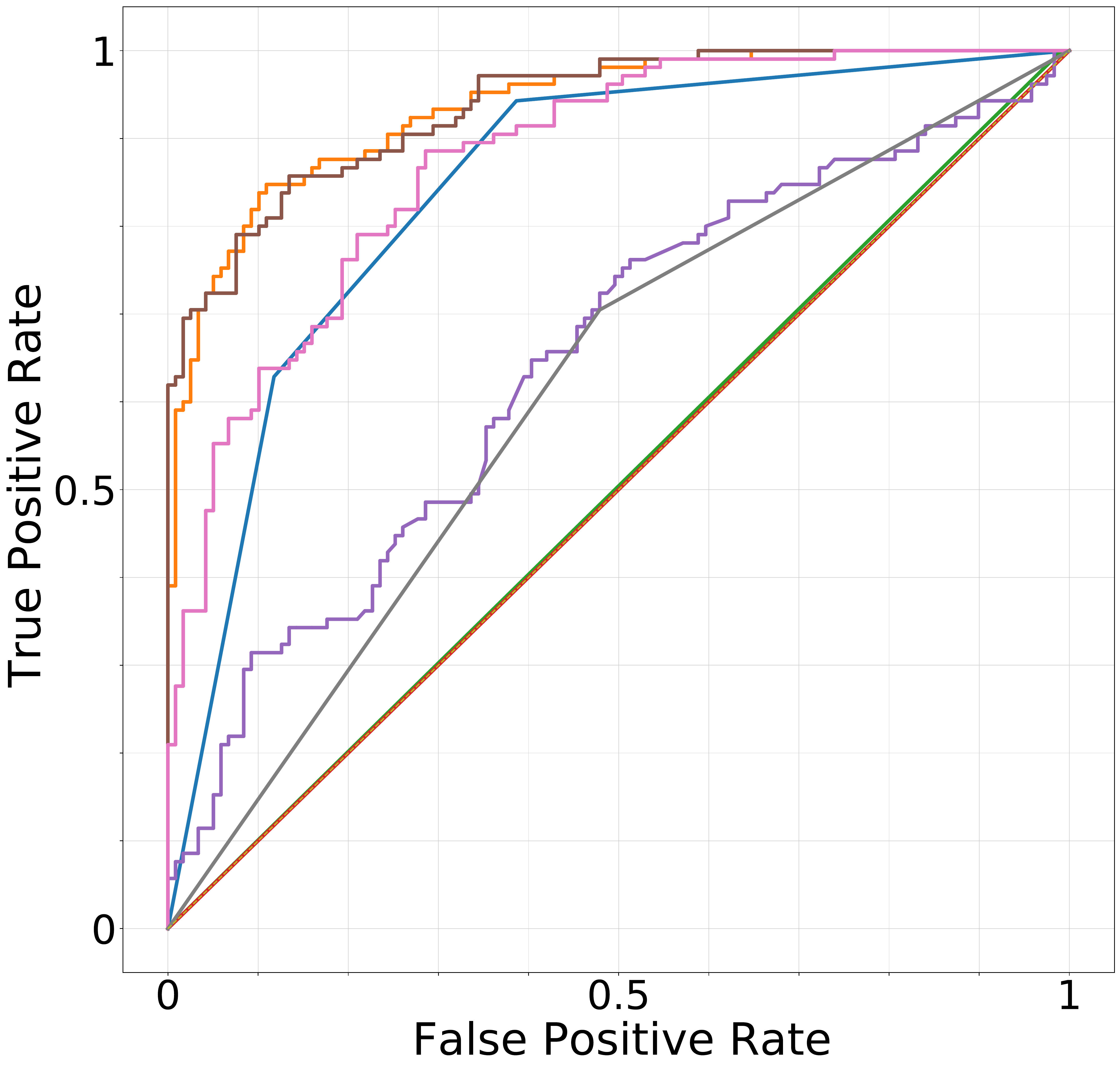}
\caption{ResNet50}
\end{subfigure}
\begin{subfigure}[b]{0.24\textwidth}
\includegraphics[width=\textwidth]{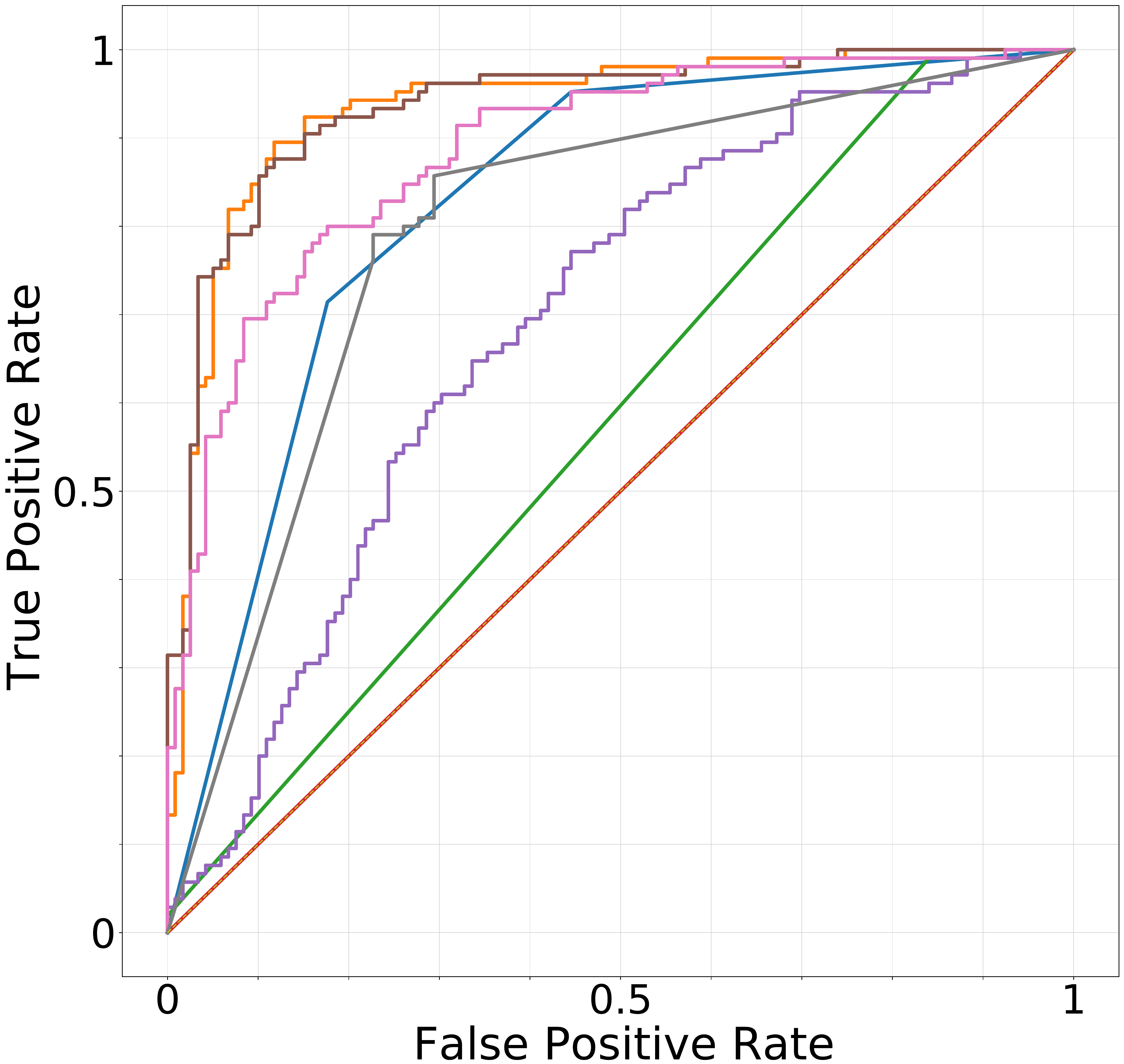}
\caption{DenseNet121}
\end{subfigure}
    \caption{ROC-AUC plots are shown for CT images using 4 architectures (VGG16, InceptionResNetV2, ResNet50, and DenseNet121) and 8 classifiers. As can be seen, linear SVM and NN (MLP) are the bests and their AUC value s are greater than others. Also it is noted that the performance of classifiers closely depends on the quality of features extracted by the convolutional layers of four considered architectures.}
    \label{fig:roc-auc}
\end{figure*}

ROC curves for all pretrained CNNs and classifiers are shown in Fig. \ref{fig:roc-auc} for a typical run. As expected, linear SVM and NN models have the highest AUC values amongst all classifiers. An important observation is that the performance of classifiers significantly varies based on hierarchically extracted features by convolutional layers of four pretrained CNNs. 

To comprehensively compare different architectures for feature extraction, we train and evaluate each classifier 100 times. Then we average all predictions to obtain a reliable estimate of the sample label. Then performance metrics including accuracy, sensitivity, specificity, and AUC value are calculated. Table \ref{tab:res-ct} and \ref{tab:res-xray} report these performance metrics for CT and X-ray datasets. Reported values are given in percent. Having compared all models, we find that no model outperforms others for most cases than others. Linear SVM also achieves best results for each model.

\begin{table}[!t]
\centering
\caption{Performance comparison of 4 architectures and 8 classifiers (32 combinations) for CT images. All values are given in percent.}\label{tab:res-ct}
\resizebox{\columnwidth}{!}{%
\begin{tabular}{@{}llllll@{}}
\toprule
Model                          & Classifier           & Accuracy & Sensitivity & Specificity & AUC  \\
\hline
\hline
\multirow{8}{*}{\rotatebox[origin=c]{90}{DenseNet121}}          & AdaBoost          & 78.9 $\pm$ 7.6    & 76.5 $\pm$ 11.5    & 80.1 $\pm$ 8.4 & 81.9 $\pm$ 5.8 \\
                                   & Gaussian Process  & 53.1 $\pm$ 10.9   & 0.0         & 1.0         & 0.5 $\pm$ 0.9 \\
                                   & Linear SVM        & \textbf{85.9} $\pm$ 5.9   & \textbf{84.9} $\pm$ 8.4      & \textbf{86.8} $\pm$ 6.3    & \textbf{93.1} $\pm$ 3.4\\
                                   & Naive Bayes       & 74.9 $\pm$ 5.5   & 72.4 $\pm$ 24.4     & 77 $\pm$ 11.5     & 77 $\pm$ 9.1 \\
                                   & Nearest Neighbors & 79.8 $\pm$ 6   & 69.7 $\pm$ 13.5      & \textbf{88.8} $\pm$ 5.3     & 86.8 $\pm$ 5.4\\
                                   & Neural Net        & \textbf{83.4} $\pm$ 14.4     & \textbf{82} $\pm$ 28      & \textbf{84.6} $\pm$ 24     & \textbf{92.7} $\pm$ 13.4 \\
                                   & RBF SVM           & 53.1 $\pm$ 2.3   & 0.0         & 1.0         & 57 $\pm$ 3.1  \\
                                   & Random Forest     & 62.3 $\pm$ 9.2    & 46.2 $\pm$ 17.7      & 76.3 $\pm$ 11.7      & 67.4 $\pm$ 10.3 \\
\hline
\multirow{8}{*}{\rotatebox[origin=c]{90}{InceptionResNetV2}}         & AdaBoost          & 75.2 $\pm$ 8.3   & 72.6 $\pm$ 12.8 & 77.5 $\pm$ 9.7      & 82.7 $\pm$ 7.4 \\
                                   & Gaussian Process  & 53.1 $\pm$ 10.9   & 0.0         & 1.0         & 0.5 $\pm$ 0.4 \\
                                   & Linear SVM        & \textbf{84.3} $\pm$ 7.3   & \textbf{83.2} $\pm$ 9 & \textbf{91.9} $\pm$ 7.4 & \textbf{91.9} $\pm$ 4.1 \\
                                   & Naive Bayes       & 75.7 $\pm$ 2.5& 76.6 $\pm$ 3.8     & 75 $\pm$ 12.8        & 77.3 $\pm$ 13.6\\
                                   & Nearest Neighbors & 75.8 $\pm$ 8.2   & 57.3 $\pm$ 14.9     & \textbf{92} $\pm$ 4.4       & 84.4 $\pm$ 6.8 \\
                                   & Neural Net        & \textbf{80.7} $\pm$ 16.5   & \textbf{79.4} $\pm$ 28.1      & 82.6 $\pm$ 29     & \textbf{87.8} $\pm$ 15.6  \\
                                   & RBF SVM           & 53.1 $\pm$ 11    & 0.0         & 1.0         & 55 $\pm$ 2.4 \\
                                   & Random Forest     & 59.1 $\pm$ 9.9   & 42.3 $\pm$ 18     & 73.4 $\pm$ 13      & 68.7 $\pm$ 11.6\\
\hline
\multirow{8}{*}{\rotatebox[origin=c]{90}{ResNet50}}          & AdaBoost          & 76.1 $\pm$ 9.4  & 73.9 $\pm$ 13.5 & 78 $\pm$ 9.8      & 83.9 $\pm$ 7.6 \\
                                   & Gaussian Process  & 53.1 $\pm$ 11    & 0.0         & 1.0         & 0.5 $\pm$ 0.4 \\
                                   & Linear SVM        & \textbf{87.9} $\pm$ 5.8    & \textbf{86.5} $\pm$ 7.1      & \textbf{89.1} $\pm$ 5.4      & \textbf{94.2} $\pm$ 2.9 \\
                                   & Naive Bayes       & 59.5 $\pm$ 9.8   & 70.1 $\pm$ 30.6      & 50 $\pm$ 25.4       & 60.1 $\pm$ 5.5 \\
                                   & Nearest Neighbors & 78.9 $\pm$ 8.2   & 65.2 $\pm$ 14.4      & \textbf{91.1} $\pm$ 4.8      & 87.2 $\pm$ 6.4 \\
                                   & Neural Net        & \textbf{84.6} $\pm$ 16.4  & \textbf{83.1} $\pm$ 30.4      & \textbf{88.3} $\pm$ 27.9      & \textbf{91.7} $\pm$ 16.7\\
                                   & RBF SVM           & 53.1 $\pm$ 11     & 0.0         & 1.0         & 51 $\pm$ 0.8  \\
                                   & Random Forest     & 59.3 $\pm$ 9.1    & 35.1 $\pm$ 16.4      & 80.6 $\pm$ 10.1      & 63.3 $\pm$ 11.4\\
\hline
\multirow{8}{*}{\rotatebox[origin=c]{90}{VGG16}}             & AdaBoost          & 76.8 $\pm$ 7.8     & 73.9 $\pm$ 13.1     & 79.2 $\pm$ 8.9      & 84.2 $\pm$ 7.1\\
                                   & Gaussian Process  & 53.1 $\pm$ 11   & 0.0         & 1.0         & 0.5 $\pm$ 0.4 \\
                                   & Linear SVM        & \textbf{86.5} $\pm$ 5.8   & \textbf{84.8} $\pm$ 8.2      & \textbf{88.1} $\pm$ 5.9      & \textbf{93.3} $\pm$ 3.3 \\
                                   & Naive Bayes       & 64.2 $\pm$ 12.2   & 74.5 $\pm$ 18.9      & 55 $\pm$ 22.5        & 64.8 $\pm$ 8.9\\
                                   & Nearest Neighbors & 71.4 $\pm$ 6.9    & 42.3 $\pm$ 14      & \textbf{97.1} $\pm$ 2.5      & 83.8 $\pm$ 6.2 \\
                                   & Neural Net        & \textbf{84.6} $\pm$ 16.3    & \textbf{85.8} $\pm$ 32.1      & \textbf{87.6} $\pm$ 27     & \textbf{92.8} $\pm$ 18.6 \\
                                   & RBF SVM           & 53.1 $\pm$ 11   & 0.0         & 1.0         & 0.5 $\pm$ 0.4 \\
                                   & Random Forest     & 59.7 $\pm$ 9   & 34.4 $\pm$ 16.4      & 82 $\pm$ 9.5        & 64.2 $\pm$ 11.8 \\ \cmidrule(l){1-6} 
\end{tabular}
}
\end{table}

\begin{table}[!t]
\centering
\caption{Performance comparison of 4 architectures and 8 classifiers (32 combinations) for X-ray images. All values are given in percent.}\label{tab:res-xray}
\resizebox{\columnwidth}{!}{%
\begin{tabular}{@{}llllll@{}}
\toprule
Model                          & Classifier           & Accuracy      & Sensitivity   & Specificity  & AUC \\
\hline
\hline
\multirow{8}{*}{\rotatebox[origin=c]{90}{DenseNet121}}      & AdaBoost          & 91 $\pm$ 5.9     & 77.7 $\pm$ 15.8 & 95.4 $\pm$ 5.2 & 96.6 $\pm$ 3.9 \\
                                   & Gaussian Process  & 74.8 $\pm$ 1.7  & 0.0           & 1.0          & 51 $\pm$ 0.8  \\
                                   & Linear SVM        & \textbf{96.4} $\pm$ 3.1  & \textbf{93.9} $\pm$ 9.3  & 97.2 $\pm$ 3.7 & \textbf{99.5} $\pm$ 0.8 \\
                                   & Naive Bayes       & 82.3 $\pm$ 5    & 31.6 $\pm$ 18.5 & \textbf{99.4} $\pm$ 0.5 & 65.5 $\pm$ 9.2 \\
                                   & Nearest Neighbors & 89.4 $\pm$ 4.6  & 64.1 $\pm$ 18.1 & 97.8 $\pm$ 2.9 & 95.4 $\pm$ 4.2 \\
                                   & Neural Net        & \textbf{93.2} $\pm$ 5.7  & \textbf{80.7} $\pm$ 24.1 & 97.4 $\pm$ 2.9 & \textbf{98.9} $\pm$ 1.2 \\
                                   & RBF SVM           & 75 $\pm$ 1.7    & 0.0           & 1.0          & 51 $\pm$ 1.1   \\
                                   & Random Forest     & 78.9 $\pm$ 4.3  & 20 $\pm$ 15.5   & \textbf{98.5} $\pm$ 2.6 & 82.3 $\pm$ 9.3 \\
\hline                                   
\multirow{8}{*}{\rotatebox[origin=c]{90}{InceptionResNetV2}} 
                                    & AdaBoost          & 89.5 $\pm$ 5.6   & \textbf{73.8} $\pm$ 17.8 & 94.8 $\pm$ 5.2 & 94.9 $\pm$ 5.4 \\
                                   & Gaussian Process  & 74.8 $\pm$ 1.7  & 0.0           & 1.0          & 51 $\pm$ 1.1   \\
                                   & Linear SVM        & \textbf{98} $\pm$ 3.2      & \textbf{96.3} $\pm$ 7.8  & 98.5 $\pm$ 3.5 & \textbf{99.8} $\pm$ 0.6 \\
                                   & Naive Bayes       & 75.3 $\pm$ 2    & 1.1           & 1.0          & 50.5 $\pm$ 1.8 \\
                                   & Nearest Neighbors & \textbf{89.9} $\pm$ 5.4  & 62.2 $\pm$ 20.3 & \textbf{99.3} $\pm$ 2.3 & 96 $\pm$ 4.7   \\
                                   & Neural Net        & \textbf{89.9} $\pm$ 12.6 & 72.7 $\pm$ 24.6 & 95.6 $\pm$ 17.1 & \textbf{97.5} $\pm$ 8.5 \\
                                   & RBF SVM           & 75.1 $\pm$ 1.7  & 0.0           & 1.0          & 51 $\pm$ 1.1   \\
                                   & Random Forest     & 77.6 $\pm$ 4.3  & 13.9 $\pm$ 13.8 & \textbf{98.9} $\pm$ 2.1 & 80.6 $\pm$ 9.7 \\
\hline                                   
\multirow{8}{*}{\rotatebox[origin=c]{90}{ResNet50}}         & AdaBoost          & 92.6 $\pm$ 6.1  & 84.4 $\pm$ 17.1 & 95.3 $\pm$ 5.5 & 97.6 $\pm$ 4.1 \\
                                   & Gaussian Process  & 75.1 $\pm$ 1.6  & 0.0           & 1.0          & 51 $\pm$ 1.1   \\
                                   & Linear SVM        & \textbf{98.6} $\pm$ 2.1  & \textbf{99.9} $\pm$ 1.2  & 98.2 $\pm$ 2.8 & \textbf{99.7} $\pm$ 0.6 \\
                                   & Naive Bayes       & 77.8 $\pm$ 3.9  & 12.5 $\pm$ 12.5 & 1.0          & 56.3 $\pm$ 6.7 \\
                                   & Nearest Neighbors & \textbf{94} $\pm$ 4.4    & 78.4 $\pm$ 17.7 & \textbf{99.2} $\pm$ 2.6 & 97.7 $\pm$ 4.9 \\
                                   & Neural Net        & 92.5 $\pm$ 6.4  & \textbf{85.5} $\pm$ 24.1 & 94.8 $\pm$ 7.3 & \textbf{98} $\pm$ 3.2     \\
                                   & RBF SVM           & 75 $\pm$ 1.7    & 0.0           & 1.0          & 51 $\pm$ 1.1     \\
                                   & Random Forest     & 76.3 $\pm$ 3.3  & 80 $\pm$ 10.8    & \textbf{99.1} $\pm$ 2 & 80.8 $\pm$ 9.6 \\
\hline                                   
\multirow{8}{*}{\rotatebox[origin=c]{90}{VGG16}}            & AdaBoost          & 89.9 $\pm$ 5.5  & 85.9 $\pm$ 17.9 & 94.9 $\pm$ 4.8 & 94.1 $\pm$ 6.8 \\
                                   & Gaussian Process  & 74.1 $\pm$ 1.7  & 0.0           & 1.0          & 51 $\pm$ 1     \\
                                   & Linear SVM        & \textbf{96.6} $\pm$ 3.4  & \textbf{98.8} $\pm$ 9.9  & 98.3 $\pm$ 3.1 & \textbf{99.6} $\pm$ 0.7 \\
                                   & Naive Bayes       & 90.2 $\pm$ 5.1  & 70.8 $\pm$ 20.3 & \textbf{99.3} $\pm$ 1.6 & 75 $\pm$ 10    \\
                                   & Nearest Neighbors & 89.3 $\pm$ 4.5  & 57.3 $\pm$ 7    & \textbf{98.9} $\pm$ 3   & 83.9 $\pm$ 8.3 \\
                                   & Neural Net        & \textbf{94.3} $\pm$ 6.1  & \textbf{98.8} $\pm$ 25.3  & 97.5 $\pm$ 5.7 & \textbf{98.3} $\pm$ 2.1 \\
                                   & RBF SVM           & 76.1 $\pm$ 1.7  & 0.0           & 1.0          & 51 $\pm$ 1     \\
                                   & Random Forest     & 76.5 $\pm$ 3.1  & 13.2 $\pm$ 10.2 & 99.5 $\pm$ 1.6 & 82.8 $\pm$ 9.3 \\ \cmidrule(l){1-6} 
\end{tabular}
}
\end{table}

Comparing the designed network to that of \cite{60}, our transfer learning-based method outperforms theirs. The best results are achieved using ResNet50 and linear SVM classifier (an accuracy of $87.9\%$). This is more than 3\% better than the best results reported in \cite{60}. ($84.7\%$ accuracy). This improvement is mainly due to a better hierarchical extraction of features using ResNet50 and an optimal selection of the classifier (linear SVM).

\begin{figure}[t]
\centering
\includegraphics[width=0.50\textwidth]{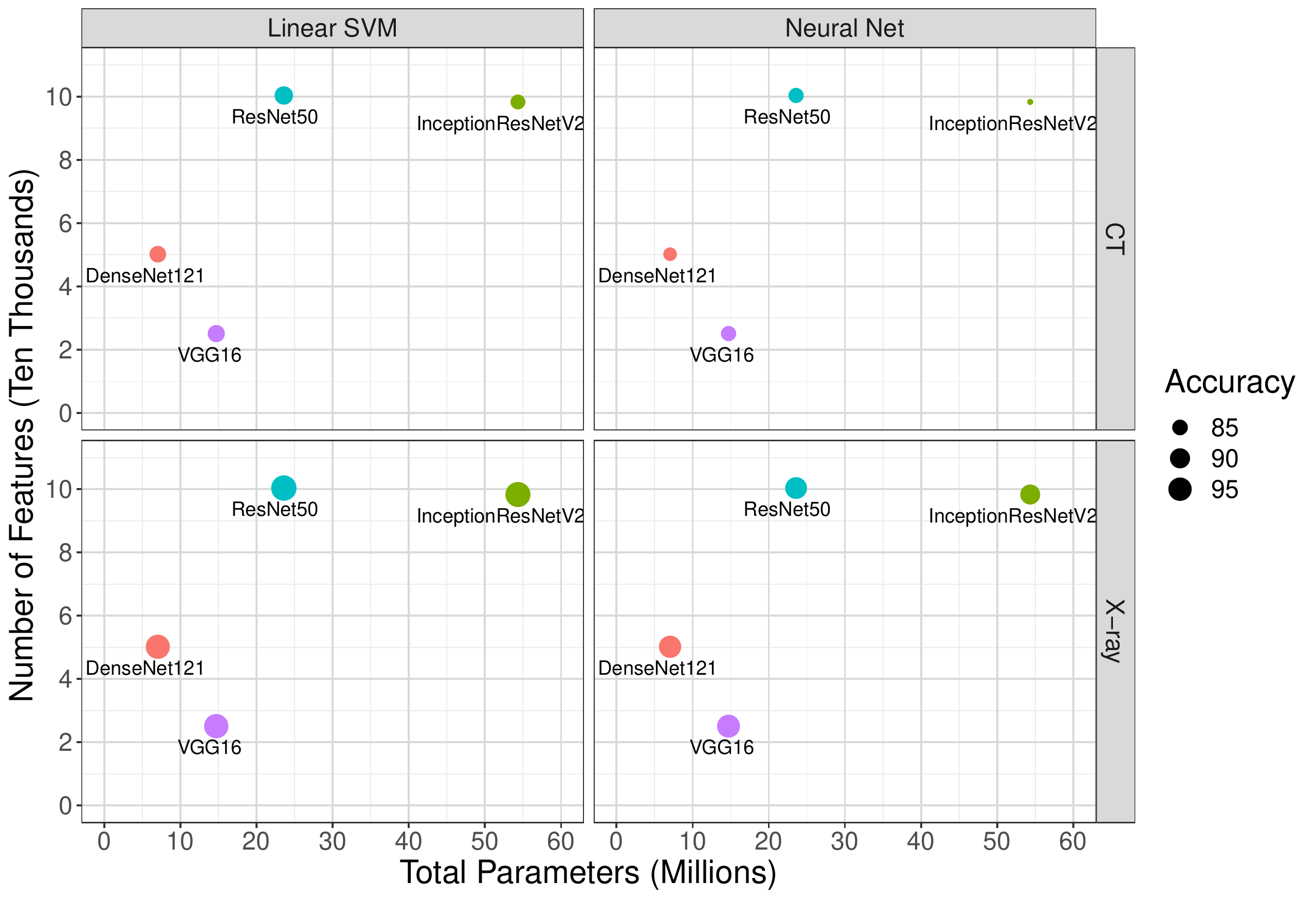}
\caption{The accuracy average in the 2D space of the number of CNN parameters (millions) and the number of features (ten thousands). The size of each point is an indication of the classifier accuracy (mean value in 100 runs).}
\label{Fig:net_size_features_acc}
\end{figure}

\begin{figure}[t]
\centering
\includegraphics[width=0.50\textwidth]{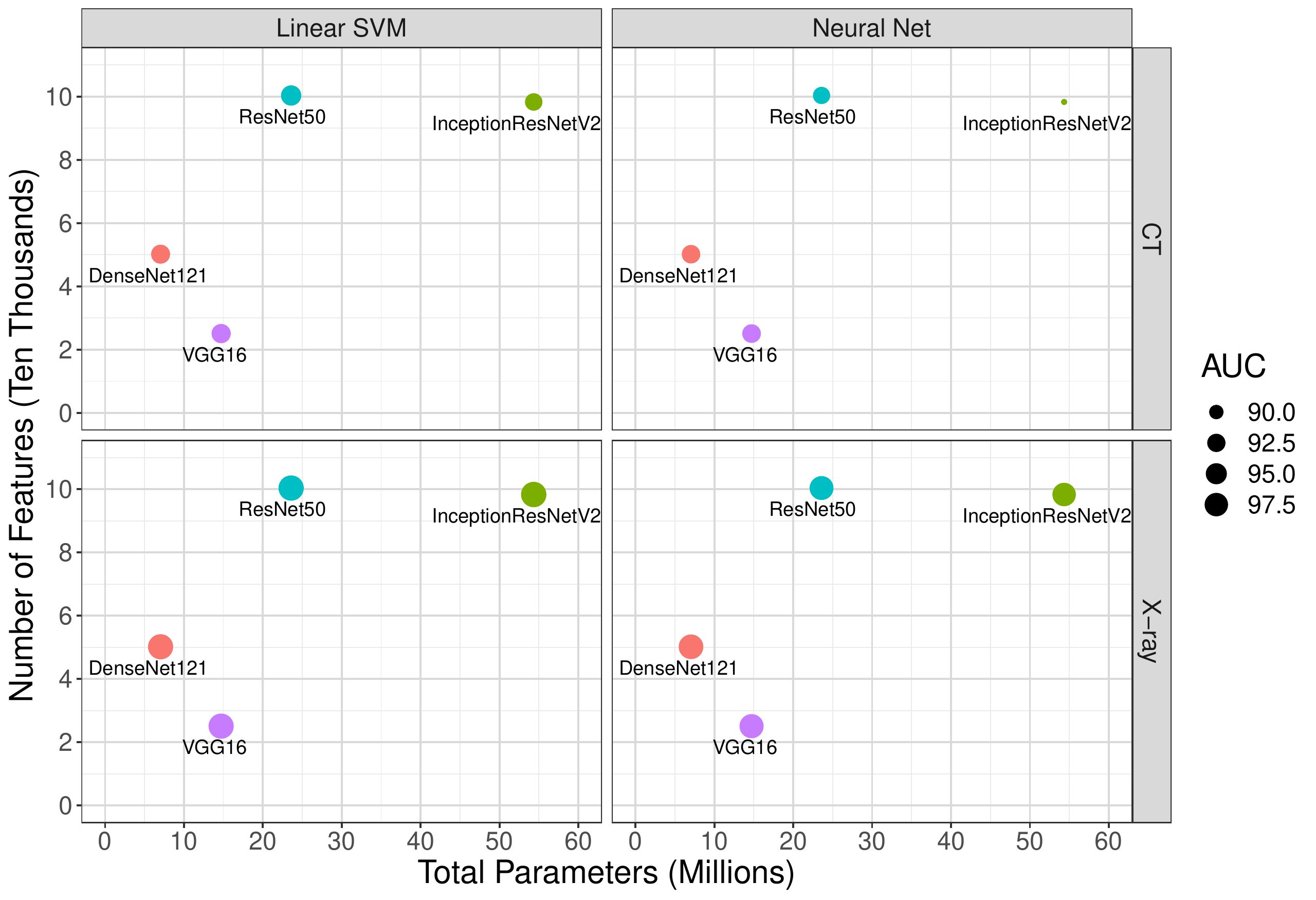}
\caption{The AUC average in the 2D space of the number of CNN parameters (millions) and the number of features (ten thousands). The size of each point is an indication of the AUC metric (mean value in 100 runs).}
\label{Fig:net_size_features_auc}
\end{figure}

It is also important to consider the network size and the number of deep features when comparing performance of pretrained CNNs for COVID-19 detection. Fig. \ref{Fig:net_size_features_acc} and Fig. \ref{Fig:net_size_features_auc} show the average of the classification performance (accuracy and AUC) for linear SVM and NN models in the 2D space formed by the number of CNN parameters (millions) and the number of features (ten thousands). The size of each point represents the accuracy and AUC metrics of the trained classifiers using features extracted by pretrained CNNs. The bigger the point, the better the performance. We generate these charts only for linear SVM and NN models as they are the best performing ones according to results presented in Table \ref{tab:res-ct} and Table \ref{tab:res-xray}. According to Fig. \ref{Fig:net_size_features_acc} and Fig. \ref{Fig:net_size_features_auc}, VGG16 is the most efficient pretrained CNNs for COVID-19 detection. It has the least number of parameters and extracts the smallest number of features. Those features are the most informative and discriminative ones as both linear and NN models achieve the best results using them. In contrast, the massive network of InceptionResNetV2 offers the most number of features that have least information content amongst investigated network. Another key observation is the choice of the pretrained CNN has a direct and profound impact on the overall performance of the COVID-19 classification model. Last but not least, one may conclude that bigger networks such as ResNet50 and InceptionResNetV2 do not necessarily extract more informative and discriminative features. 

\begin{figure*}[t]
\centering
\begin{subfigure}[b]{0.24\textwidth}
\includegraphics[width=\textwidth]{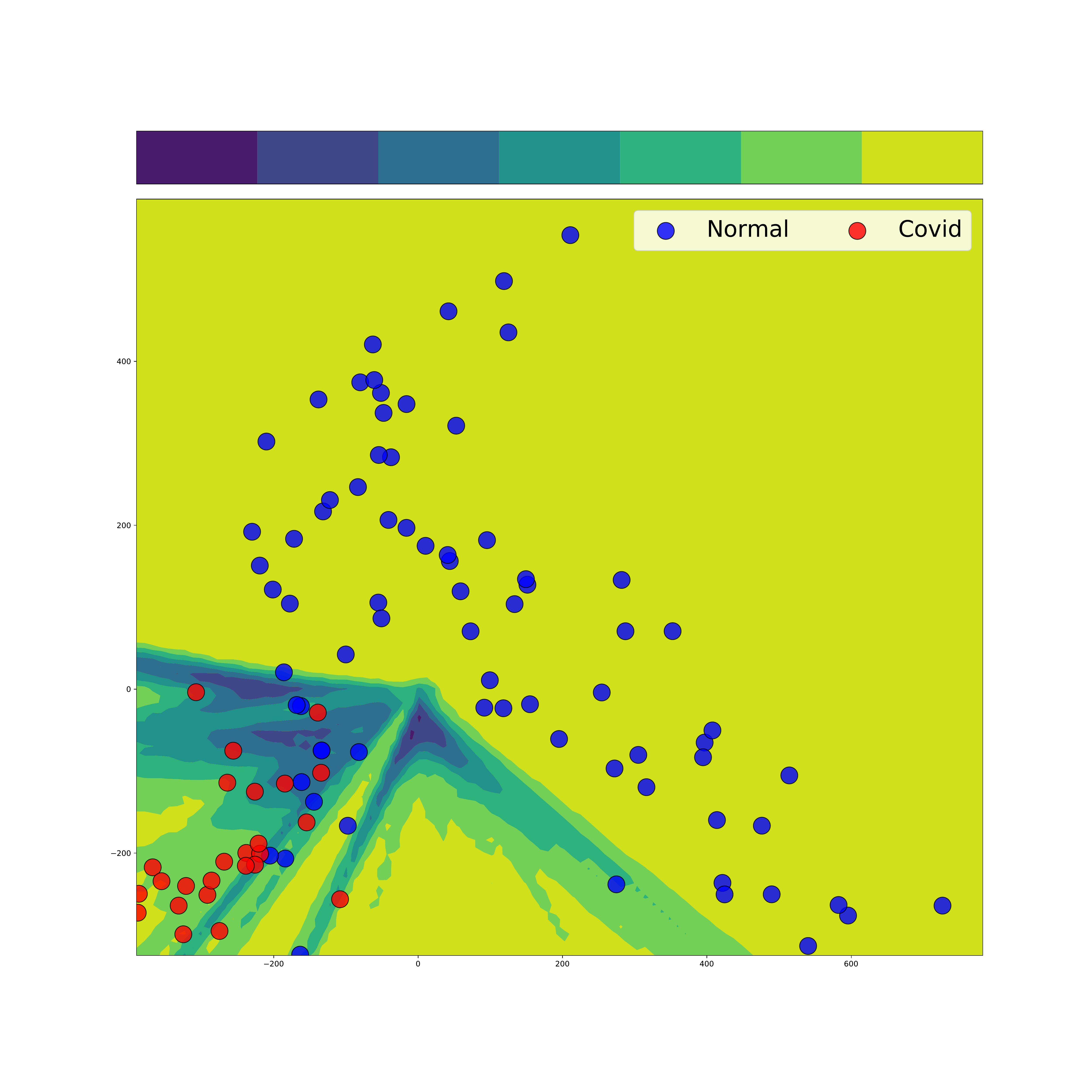}
\caption{VGG16}
\label{Fig:res-unc-xray-VGG16}
\end{subfigure}
\begin{subfigure}[b]{0.24\textwidth}
\includegraphics[width=\textwidth]{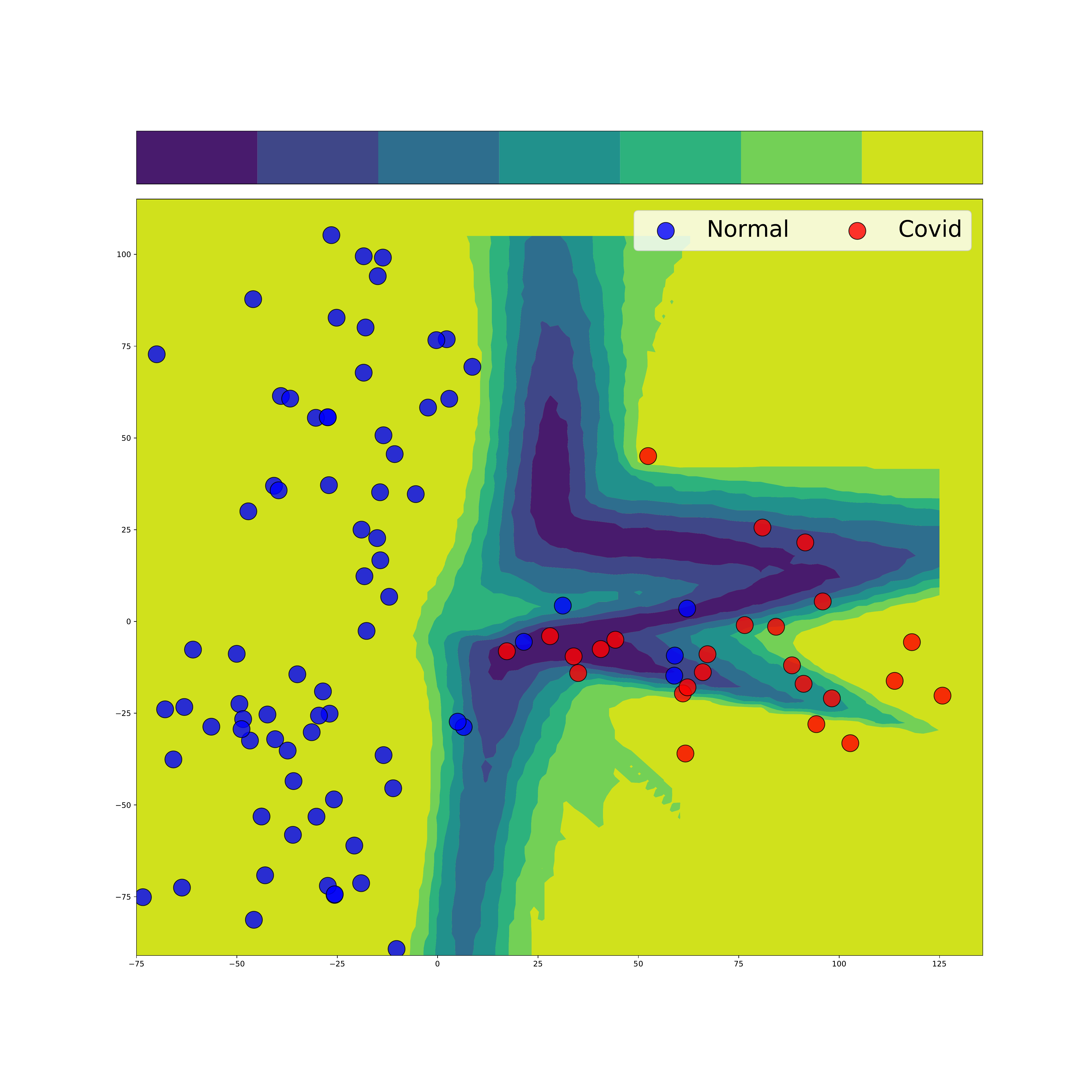}
\caption{InceptionResNetV2}
\end{subfigure}
\begin{subfigure}[b]{0.24\textwidth}
\includegraphics[width=\textwidth]{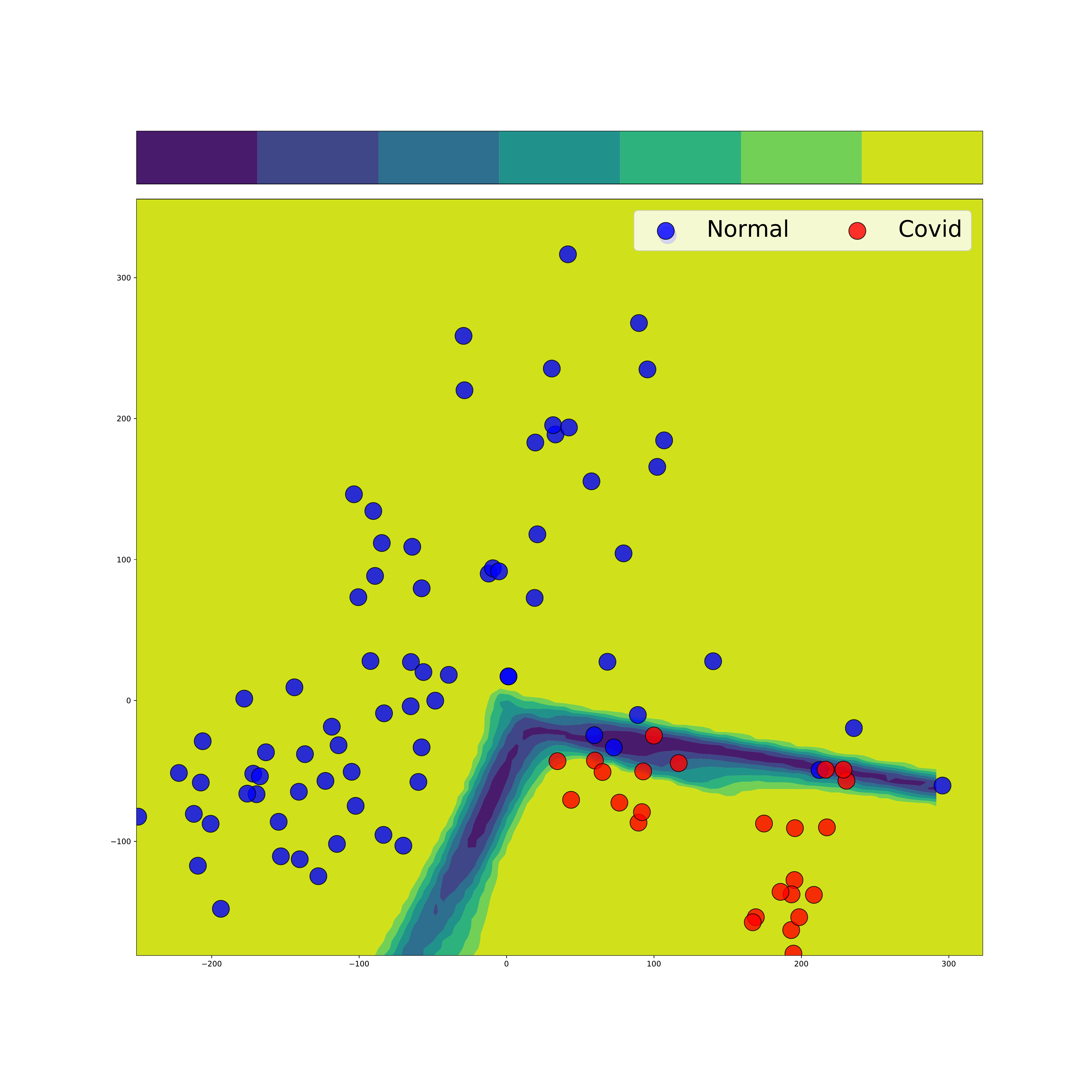}
\caption{ResNet50}
\end{subfigure}
\begin{subfigure}[b]{0.24\textwidth}
\includegraphics[width=\textwidth]{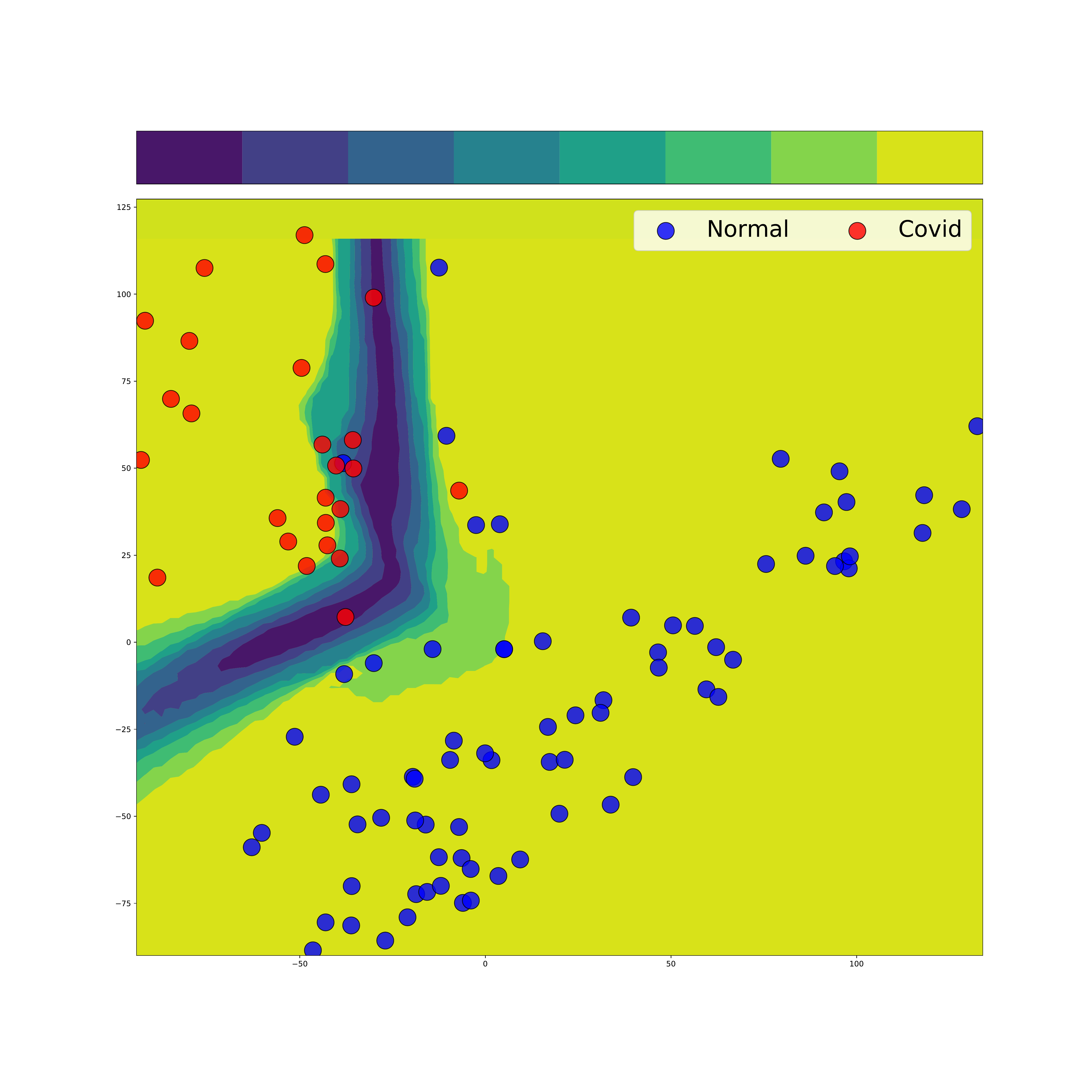}
\caption{DenseNet121}
\label{Fig:res-unc-xray-DenseNet121}
\end{subfigure}
\begin{subfigure}[b]{0.24\textwidth}
\includegraphics[width=\textwidth]{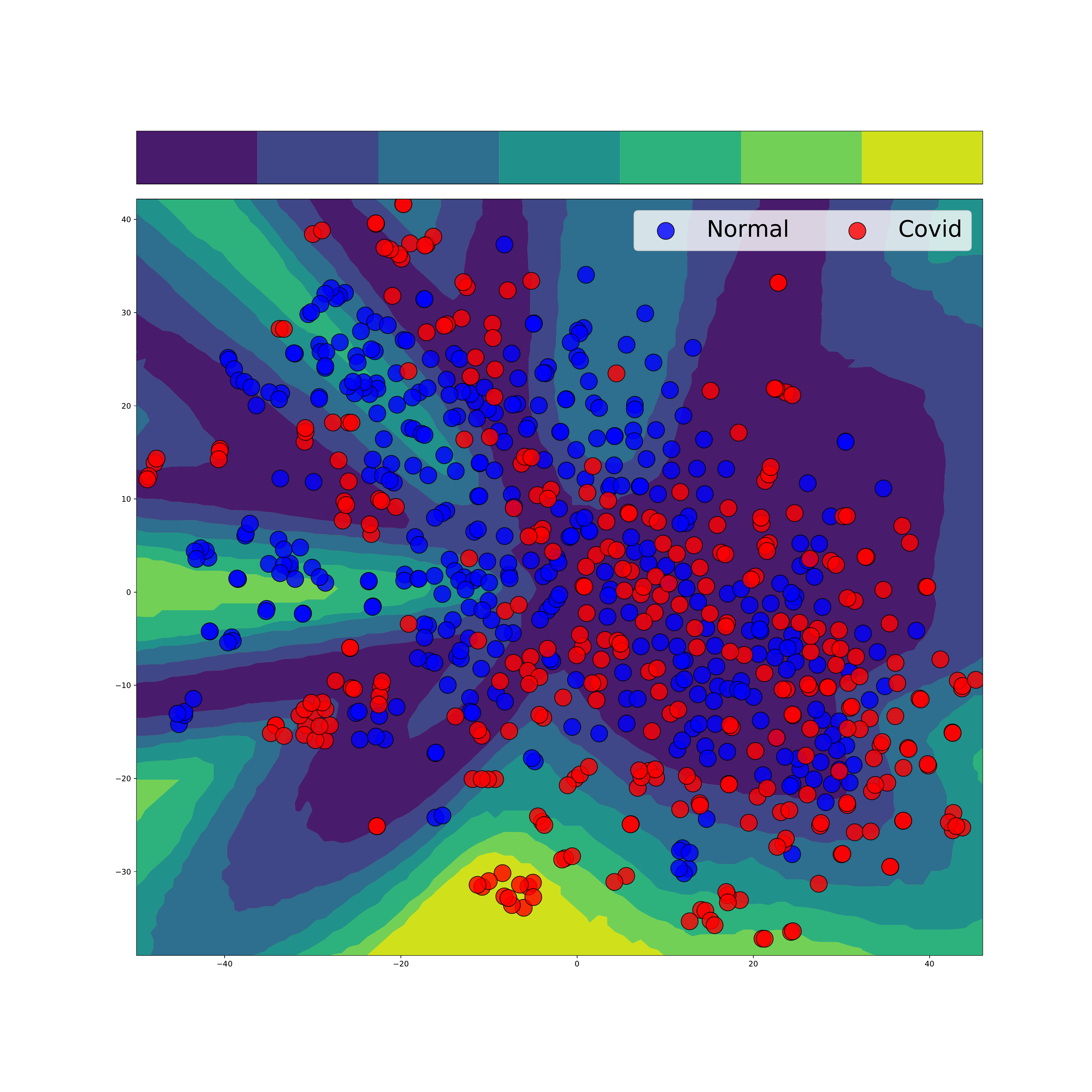}
\caption{VGG16}
\label{Fig:res-unc-ct-VGG16}
\end{subfigure}
\begin{subfigure}[b]{0.24\textwidth}
\includegraphics[width=\textwidth]{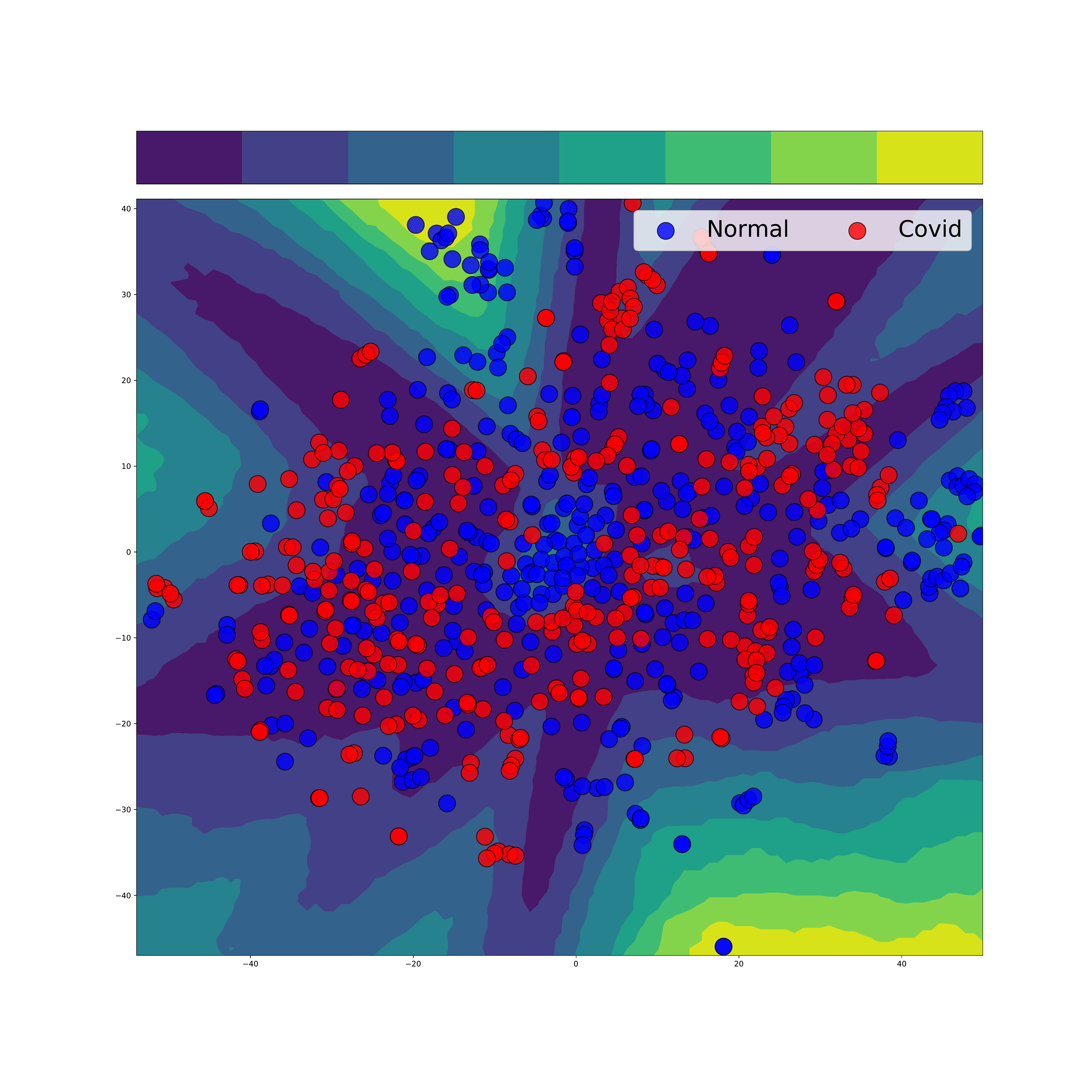}
\caption{InceptionResNetV2}
\end{subfigure}
\begin{subfigure}[b]{0.24\textwidth}
\includegraphics[width=\textwidth]{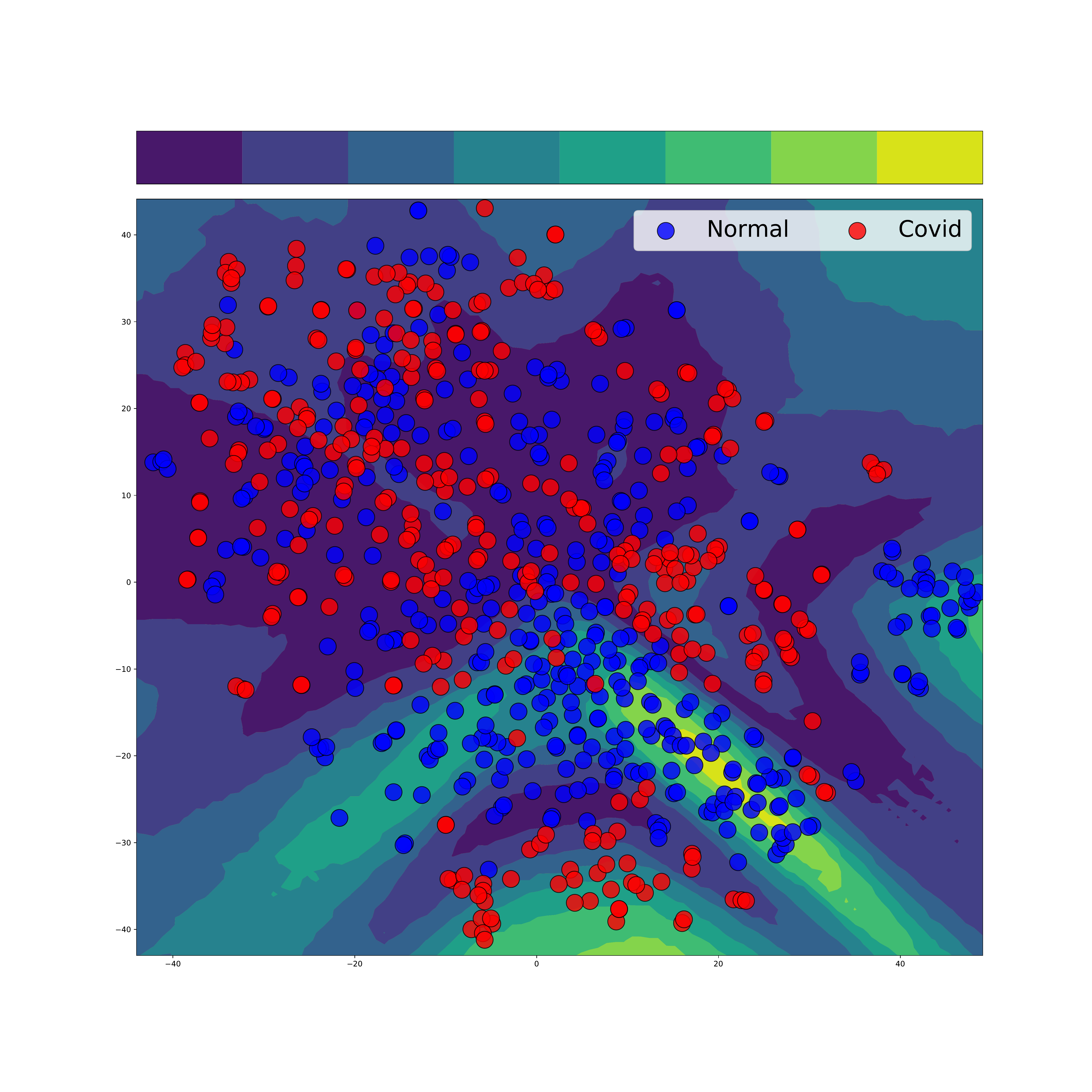}
\caption{ResNet50}
\end{subfigure}
\begin{subfigure}[b]{0.24\textwidth}
\includegraphics[width=\textwidth]{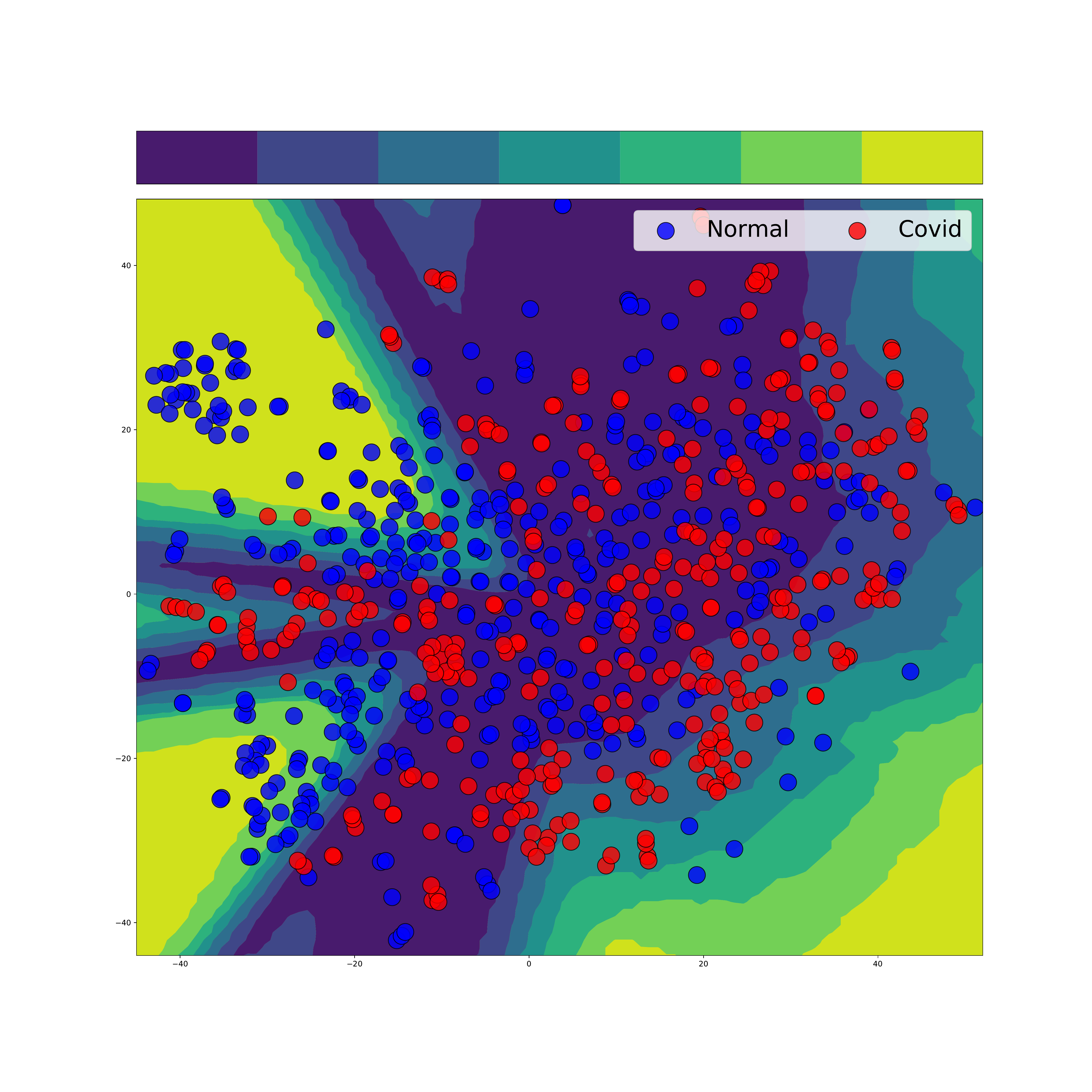}
\caption{DenseNet121}
\label{Fig:res-unc-ct-DenseNet121}
\end{subfigure}

    \caption{Uncertainty quantification using $20$ individual neural networks working as an ensemble. They differ in the number of neurons in their hidden layer before applying multi-layer perceptron classifier. The darker the color, the higher the uncertainty level. Samples on dark parts of the plot have a high level of predictive uncertainty as the 20 models could not all agree on the predicted label..}
    \label{Fig:res-uncertainty}
\end{figure*}

%
%
%

It is also quite important to quantify uncertainties associated with predictions. Here, we generate predictive uncertainty estimates for NN models. As mentioned before, there are several ways of generating ensemble networks. We use the entire dataset in the training step because the availability of more samples improves the generalization power of NN models. Twenty individual NN models with different architectures are first trained to form an ensemble. NN models have a hidden layer and their number of hidden neurons is randomly selected between 50 to 400. 

Fig. \ref{Fig:res-uncertainty} shows the predictive uncertainty estimates for both Covid and non-Covid cases in a 2D space. Fig. \ref{Fig:res-unc-xray-VGG16}-\ref{Fig:res-unc-xray-DenseNet121} display epistemic uncertainties for the X-ray dataset. Fig. \ref{Fig:res-unc-ct-VGG16}-\ref{Fig:res-unc-ct-DenseNet121} show epistemic uncertainties for the CT dataset. The 2D space is obtained after applying PCA to reduce dimensionality of obtained features from pretrained CNNs. This projection to the 2D space is done to ensure that samples could be visualized against calculated predictive uncertainty estimates. The darker the color of filled area, the higher the uncertainty level. According Fig. \ref{Fig:res-unc-xray-VGG16}-\ref{Fig:res-unc-xray-DenseNet121}, the level of epistemic uncertainty for the X-ray dataset is fairly low. While the projected features in the 2D space are in different locations for four pretrained CNNs, the NN classifiers generate very similar results. This consistency leads to a low uncertainty. In contrast, the predictive uncertainty estimates for CT images are quite high. This is evident from the large dark area in Fig. \ref{Fig:res-unc-ct-VGG16}-\ref{Fig:res-unc-ct-DenseNet121}. These indicate that individual NN models in the ensemble have a different generalization power and produce significantly inconsistent results. There is no perfect agreement between all models about the predictive labels of these samples. Accordingly, more care should be exercised when using machine learning predictions for COVID-19 diagnosis using CT images.

\section{Conclusion}\label{Sec:concl}
The purpose of this study was to investigate the suitability of deep transfer learning for COVID-19 diagnosis using medical imaging. The key motivation was the lack of access to large repositories of images for developing deep neural networks from scratch. Leveraging the transfer learning framework, we apply $4$ pretrained deep convolutional neural networks (VGG16, ResNet50, DenseNet121, and InceptionResNetV2) to hierarchically extract informative and discriminative features from chest X-ray and CT images. The parameters of the convolutional layers are kept frozen during the training process. Extracted features are then processed by multiple classification techniques. Obtained results indicate that linear SVM and multi-layer perceptron outperforms others methods in terms of the medical diagnosis accuracy for both X-ray and CT images. It is also observed that better prediction results and medical diagnosis could be achieved using CT images as they are much richer in information compared to X-ray images. 

There are many rooms for improvement and further exploration. The performance of transfer learning algorithms could be majorly improved by fine tuning them to extract more informative and discriminative features. Features obtained from different transfer learning models could be combined to develop hybrid models. Also, predictions from individual models could be combined to form ensembles. Last but not least, state of the art method could be applied for more comprehensive estimation of the uncertainty measures.

\ifCLASSOPTIONcaptionsoff
  \newpage
\fi


\bibliographystyle{ieeetr}

\end{document}